\documentclass[fleqn,usenatbib]{mnras}

\usepackage{newtxtext,newtxmath}

\usepackage[T1]{fontenc}

\DeclareRobustCommand{\VAN}[3]{#2}
\let\VANthebibliography\thebibliography
\def\thebibliography{\DeclareRobustCommand{\VAN}[3]{##3}\VANthebibliography}

\usepackage{graphicx}	
\usepackage{amsmath}	
\usepackage{bbm} 
\usepackage{verbatim}
\usepackage{url}
\usepackage{aas_macros}
\usepackage{siunitx} 
\usepackage{pdflscape} 

\defcitealias{2013MNRAS.430..174H}{Paper~1}
\defcitealias{2014MNRAS.443.1482H}{Paper~2}
\defcitealias{2016MNRAS.461.2025E}{Paper~3}
\defcitealias{2019MNRAS.490.5807E}{Paper~4}

\title[Numerical modelling of radio galaxy lobes - realistic clusters]{Numerical modelling of the lobes of radio galaxies - Paper V: Universal Pressure Profile cluster atmospheres}

\author[M. Stimpson et al.]{
M. Stimpson,$^{1}$\thanks{E-mail: mstimpson2@herts.ac.uk}
M.J. Hardcastle,$^{1}$
M. G. H. Krause$^{1}$
\\
$^{1}$Centre for Astrophysics Research, Department of Physics, Astronomy and Mathematics, University of Hertfordshire, College Lane, Hatfield, Hertfordshire AL10 9AB, UK\\
}

\date{Accepted XXX. Received YYY; in original form ZZZ}

\pubyear{2022}

\begin{document}
\label{firstpage}
\pagerange{\pageref{firstpage}--\pageref{lastpage}}
\maketitle

\begin{abstract}
We present relativistic magnetohydrodynamic modelling of jets running into hydrostatic, spherically symmetric cluster atmospheres.  For the first time in a numerical simulation, we present model cluster atmospheres based upon the Universal Pressure Profile (UPP), incorporating a temperature profile for a `typical' self-similar atmosphere described by only one parameter - $M_{500}$.  We explore a comprehensive range of realistic atmospheres and jet powers and derive dynamic, energetic and polarimetric data which provide insight into what we should expect of future high-resolution studies of AGN outflows.  From the simulated synchrotron emission maps which include Doppler beaming we find sidedness distributions that agree well with observations.  We replicated a number of findings from our previous work, such as higher power jets inflating larger aspect-ratio lobes and the cluster environment impacting the distribution of energy between the lobe and shocked regions.  Comparing UPP and $\beta$-profiles we find that the cluster model chosen results in a different morphology for the resultant lobes with the UPP more able to clear lobe material from the core; and that these different atmospheres influence the ratio between the various forms of energy in the fully developed lobes.  This work also highlights the key role played by Kelvin-Helmholtz (KH) instabilities in the formation of realistic lobe aspect-ratios.  Our simulations point to the need for additional lobe-widening mechanisms at high jet powers, for example jet precession.  Given that the UPP is our most representative general cluster atmosphere, these numerical simulations represent the most realistic models yet for spherically symmetric atmospheres.
\end{abstract}




\section{Introduction}\label{intro}
The modelling of the lobes of AGN jets began with analytical approaches such as the key work of \cite{1974MNRAS.166..513S} and later on \cite{1997MNRAS.286..215K}.  Access to high-performance computing in recent years has fuelled modelling using numerical simulations:  varying the input parameters of our models and comparing the output with observations tells us much about the conditions and processes in the AGN.  Early simulations, such as the 2DHD model of \cite{1982A&A...113..285N} confirmed the predictions made by \cite{1974MNRAS.169..395B} about bipolar jets creating hot spots and a bow shock at the jet head, followed by a contact discontinuity between the lobe and shocked ambient medium.  More recent models incorporate more of the physics of actual AGN's, such as the 3DRMHD work of \cite{2010MNRAS.402....7M} who find that 3D models reveal kink instabilities which are not seen in similar 2D models.  Recent reviews of numerical modelling of jets include \cite{2019Galax...7...24M}, \cite{2021NewAR..9201610K} and \cite{2023Galax..11...73B}.  A key role for modelling now is to predict the images which will be captured by the next generation of X-ray satellites such as \textit{Athena} \citep{2013arXiv1306.2307N}, and of radio telescopes such as the \textit{Square Kilometer Array} \citep{2004NewAR..48..979C}.

The primary factor which shapes large-scale jet structure is the power of the jet \citep{1991Natur.349..138R, 2009A&ARv..17....1W, 2021A&ARv..29....3O}.  Using a method based upon observations of the jet terminal hotspots, \cite{2013ApJ...767...12G} obtained jet powers for FR IIs in the range \num{e38} to \num{e39}W; in good agreement with other researchers using a variety of techniques \citep{2014Natur.515..376G, 2012AdAst2012E...6G}.  \cite{1997MNRAS.286..215K} found that greater powers ($Q > \num{e37}$W) formed FR IIs; similar relations have been demonstrated by the simulations of \cite{2016A&A...596A..12M} and \cite{2018MNRAS.481.2878E}.  Furthermore, relativistic MHD simulations conclude that at the intermediate power between FR I and FR II, it is other factors such as the density ratio, Lorentz factor and magnetic field strength which determines whether the lobe is FR I or FR II \citep{2010MNRAS.402....7M, 2020MNRAS.499..681M, 2022A&A...659A.139M}. \cite{2022A&A...659A.139M} note that the observation epoch also plays a role as many features are time-dependent. Perpendicular shocks close to the radio core have been shown to form in jets with wide initial opening angle \citep{2012MNRAS.427.3196K, 2022arXiv221210059Y}. Particle acceleration at such shocks could be responsible for the brightening of FR I jets at flaring points.

The structure of the cluster atmosphere is also a key factor which will influence the morphology of the lobes.  Early researchers used uniform distributions \citep{1982A&A...113..285N, 1988A&A...206..204K, 1989ApJ...344...89L} which could only represent real atmospheres over short distances; an improvement on these is the now widely-used and more realistic $\beta$-model \citep{2002MNRAS.332..271R, 2003MNRAS.339..353B, 2003A&A...402..949Z, 2005A&A...431...45K}.  Observations of powerful radio sources has demonstrated that asymmetries in the distribution of ionized gas is correlated with the structural asymmetry of the radio lobes \citep{1989AJ.....98.1232P, 1991ApJ...371..478M, 2000A&A...363..507G}, which indicates that environmental asymmetries play a role in creating structural asymmetries in the radio lobes; as demonstrated in numerical models of jets propagating through inhomogeneous environments \citep[e.g.][]{2005A&A...432..823J, 2009MNRAS.400.1785G, 2011MNRAS.411..155G, 2021MNRAS.508.5239Y,2022MNRAS.511.5225Y}; in particular, \cite{2022AJ....163..134T} demonstrated that lower power jets are impacted more and \cite{2011ApJ...728...29W} and \cite{2012ApJ...757..136W} showed that inhomogeneities in the ICM impact the transfer of energy from the lobe to the surroundings.  Furthermore, using an environment derived from a simulation of a dynamically active cluster represents another step towards greater realism \citep[e.g.][]{2006MNRAS.373L..65H, 2012ApJ...750..166M}.  Such large scale motion of the ICM disrupts the jets and plays a significant role in spreading out the energy injected \citep{2010MNRAS.407.1277M, 2017MNRAS.472.4707B, 2019MNRAS.490..343B, 2021MNRAS.506..488B}.  Dynamic modelling has also been used to investigate self-regulated feedback \citep{2017ApJ...841..133M, 2023MNRAS.518.4622E}, accretion rates \citep{2015ApJ...811..108P} and energy transfer mechanisms \cite{2018MNRAS.481.2878E}.  For reviews see \cite{2017ApJ...841..133M} and \cite{2023Galax..11...73B}.  In this work we model a spherically symmetric hydrostatic profile; similar models have been used by others to investigate feedback mechanisms and energy transfer \citep[e.g.][]{2016ApJ...829...90Y, 2016ApJ...818..181Y, 2017MNRAS.470.4530W, 2023MNRAS.523.1104W}.  However, for the first time in a numerical simulation, we use the Universal Pressure Profile \citep{2010A&A...517A..92A} as a generalised hydrostatic, spherically symmetric cluster atmosphere.

The numerical model used in this study is a development of that described in \cite{2013MNRAS.430..174H}, \cite{2014MNRAS.443.1482H}, \cite{2016MNRAS.461.2025E} and \cite{2019MNRAS.490.5807E} (henceforth referred to as \citetalias{2013MNRAS.430..174H,2014MNRAS.443.1482H,2016MNRAS.461.2025E} and \citetalias{2019MNRAS.490.5807E}).  In all these previous papers we used a $\beta$-atmosphere, but here we use the more realistic UPP atmosphere, as well as a jet with a higher Lorentz factor ($\gamma = 10$) which matches well with the values seen on parsec scales.  The model also employs stretched grids in order to model the central regions at a much higher resolution and so enable a considerably narrower, and so more realistic, injection cylinder.  In Section 2 we will present the UPP atmosphere used in this study, including the temperature profile and the ambient magnetic field.  In Section 3 we describe how the simulated atmosphere is implemented and the jet parameters used.  Our results are presented in Section 4; in Section 5 we discuss the findings and our Summary and Conclusions are found in Section 6.

\section{Cluster Atmospheres}

\subsection{Galaxy cluster density and pressure distributions}\label{clusteratmospheres}
 The X-ray emission from the hot plasma in galaxy clusters was studied by \cite{1976A&A....49..137C} who developed a model in which the material of the cluster has a density profile described by King's approximation \citep{1962AJ.....67..471K}.  This self-consistent isothermal model is referred to as the $\beta$-model and is widely used to describe the density profile in clusters of galaxies, it can be written
\begin{equation}
\rho(r)=\rho_0\left[1+ \left( \frac{r}{r_c} \right)^2 \right]^{-\frac{3\beta}{2}}\label{beta}
\end{equation}
where $\rho_0$ is the density at the centre of the cluster, $r_c$ is the `core radius' and $\beta$ is an indication of the gradient beyond the core radius.  Despite its wide use in simulations, it has long been recognised that the $\beta$-model does not adhere well to observations, particularly near the core where observations indicate a `cusp' rather than the constant density produced by the beta model \citep{1988ApJ...327..507F}.

The Navarro, Frenk and White (NFW) model also describes the density profile of the cluster and is an improvement on the $\beta$-model.  The authors describe the NFW as a `universal' density profile; it was the result of conducting a series of N-body simulations of higher resolution than previous studies \citep{1995MNRAS.275..720N}. This removed the constant density of the $\beta$-model and replaced it with more of a `cusp'.

Based upon observations of X-ray clusters with \textit{Chandra} and on numerical simulations on scales larger than these, \cite{2007ApJ...668....1N} proposed a `generalised' NFW (GNFW) model in which they parametrized the profile further; instead of expressing their distribution in terms of density (as the $\beta$ and the NFW models were) they expressed it in terms of the gas pressure of the cluster.  The version below is that presented by \cite{2010A&A...517A..92A} and is written in terms of the average scaled pressure $\mathbbm{p}$ at a normalized distance $x$ (Equation \ref{UPP2}) from the cluster centre, the profile is
\begin{equation}
\mathbbm{p}(x)=\frac{P_0}{\left(c_{500}x\right)^{\gamma}\left[1+\left(c_{500}x\right)^\alpha\right]^{\left(\beta-\gamma\right)/\alpha}},\label{UPP}
\end{equation}
where $P_0$ is the pressure at the centre of the cluster and the parameters $\gamma,\alpha, \beta$ are respectively the central slope $\left(r \ll r_s\right)$, intermediate slope $\left(r \sim r_s \right)$ and outer slope $\left(r \gg r_s\right)$.  The scale radius $r_s$, is defined as the radius where the logarithmic slope of the density profile is $\alpha=-2$; and the concentration is defined as $c_{500}\equiv R_{500}/r_s$.  $R_{500}$ represents the radius of the cluster corresponding to a mean mass density contrast of $500$ times the critical density of the Universe.  These parametrized values are linked to real values of pressure $P(r)$ and radial distance $r$ using the scaling relations
\begin{equation}
P(r)=p(x)P_{500} \qquad \text{and} \qquad x \equiv r/R_{500}\label{UPP2}
\end{equation}
where $p(x)$ is the normalized pressure and is linked to the average scaled profile $\mathbbm{p}(x)$ by an empirical term which reflects the deviation from standard self-similar scaling:
\begin{equation}
p(x)=\mathbbm{p}(x) \left[ \frac{M_{500}}{\num{3e14} \text{h}^{-1}_{70}M_{\odot}} \right]^{\alpha(x)}\label{mass_dep}
\end{equation}
where $\alpha(x)$ is a variable in $x$ linked to the mass of the cluster and the dimensionless Hubble constant h$_{70}=$ h/H$_0$ where H$_0=70\,$kms$^{-1}$Mpc$^{-1}$.  $P_{500}$ is the `characteristic pressure' which is dependent upon mass and redshift as follows
\begin{equation}
P_{500}=\num{1.65e-3}h(z)^{8/3} \left[ \frac{M_{500}}{\num{3e14} \text{h}^{-1}_{70}M_{\odot}} \right]^{2/3} \text{h}^2_{70}\,\text{keV}\,\text{cm}^{-3}\label{UPP3}
\end{equation}
where $h(z)$ is the ratio of the Hubble constant at redshift $z$ to its present value, $h(z)=H(z)/H_0$.  $M_{500}$ is the mass contained within the radius $R_{500}$ at which the mean mass density is $500$ times that of the critical density of the Universe at the cluster redshift $\rho_c(z)$.  $M_{500}$ and $R_{500}$ can be found from one another; from the definition of $M_{500}$ we have
\begin{equation}
M_{500}=\frac{4\pi}{3}R^3_{500}500\rho_c(z)\qquad \text{where} \qquad \rho_c(z)=\frac{3H(z)^2}{8\pi G}\label{M500R500}
\end{equation}
where $G$ is the gravitational constant and $H(z)=H_0\sqrt{\Omega_M(1+z)^3+\Omega_{\Lambda}}$ where, for a flat $\Lambda$CDM cosmology, $\Omega_M=0.3$ and $\Omega_{\Lambda}=0.7$.  These relations are the GNFW model and the correct choice of parameters will result in a very good fit to the pressure profiles of galaxy clusters (as shown in Appendix C of \cite{2010A&A...517A..92A}).  From the UPP cluster pressure profile described here and using a cluster temperature profile (see Section \ref{temperature_variation}) we can recover the density profile using the ideal gas law $p=nkT$.  A comparison between the $\beta$ and the UPP density profiles can be seen in Figure \ref{BETAUPPgraph}.

\subsection{The Universal Pressure Profile}\label{UPP_section}

\begin{table}
\caption{UPP parameters.  The data used in \protect\cite{2010A&A...517A..92A} is based upon 33 local ($z<0.2$) clusters drawn from the REFLEX catalogue and observed with \textit{XMM-Newton} with mass in the range $\num{e14}M_{\odot}<\num{e15}M_{\odot}$.  The data used by \protect\cite{2021ApJ...908...91H} is derived from simulations compared with X-ray data (REXCESS).}
\label{UPP_parameters}
\centering
\begin{tabular}{cccccc}
\hline
 References & $P_0$ & $c_{500}$ & $\gamma$ & $\alpha$ & $\beta$\\
 \hline
 {\cite{2010A&A...517A..92A}} & 8.130$\text{h}_{70}^{-3/2}$ & 1.156 & 0.3292 & 1.0620 & 5.4807\\
 {\cite{2021ApJ...908...91H}} & 5.048$\text{h}_{70}^{-3/2}$ & 1.217 & 0.433 & 1.192 & 5.490\\
 \hline
\end{tabular}

\end{table}

\cite{2010A&A...517A..92A} derived an average GNFW profile (see Table \ref{UPP_parameters} for the parameter values).  For their choice of parameters the scaled pressure profiles do not show any significant dependence on mass; in other words Equation \ref{mass_dep} reduces to $p(x)=\mathbbm{p}(x)$ and so their model is self-similar.

When X-ray measurements are used to estimate cluster masses, it is assumed that the cluster is in equilibrium (i.e. a perfectly relaxed cluster) \cite[e.g.][]{2010A&A...517A..92A}; however, observations indicate that clusters are not all relaxed as non-thermal pressure support is also present \cite[e.g.][]{2004A&A...426..387S, 2011MNRAS.410.1797S,  2015MNRAS.453.3699W, 2016Natur.535..117H, 2016A&A...585A.130H, 2019A&A...621A..40E, 2018ApJ...861...71S}; furthermore, simulations of clusters undergoing mergers or feedback processes substantiate these observations \citep[e.g][]{2012A&A...544A.103V, 2017MNRAS.464..210V, 2014ApJ...782..107N, 2017MNRAS.467.3737G, 2022MNRAS.514..313B} demonstrating that cluster formation leads to significant non-thermal gas processes such as turbulent flows and bulk motions.  Neglecting the kinematics leads to a systematic underestimation of the masses of galaxy clusters: this is the hydrostatic mass bias. \cite{2021ApJ...908...91H} describe how they employ a simulation (the Mock-X analysis framework) devised by \cite{2021MNRAS.506.2533B} which is able to model the evolution of clusters, including the non-thermal pressure support, and simulate X-ray emission.  Their study leads to the debiased values for the GNFW parameters (see Table \ref{UPP_parameters}) which can be considered to be more accurate than those provided by previous studies; in addition, they confirmed the self-similarity conclusion of \cite{2010A&A...517A..92A} in that this set of generalised parameters does not depend upon mass.  Their findings indicate that the UPP of \cite{2010A&A...517A..92A} is $\approx 5\%$ higher than their debiased pressure values at the centre, rises to $\approx 20\%$ at $R_{500}$ and reaches almost $\approx 35\%$ in the outermost regions.  We will use the debiased UPP values of \cite{2021ApJ...908...91H} in this study.

\subsection{Cluster temperature profile and magnetic fields}\label{temperature_variation}
In a \textit{cool core} cluster the temperature rises steeply away from the centre and reaches a peak around a tenth of $R_{500}$, then reduces gradually towards large radii.  The core is believed to be at a lower temperature as a result of radiative cooling; the inner regions are at a higher pressure than the $\beta$-model predicts and this leads to a greater luminosity and so shorter cooling time than the rest of the cluster.  Without compensation through a heating mechanism, the core temperature falls \citep{1994ARA&A..32..277F}.  \cite{2006ApJ...640..691V}, using \textit{Chandra} data for a sample of nearby relaxed clusters, derived the following radial variation of temperature $T(r)$:
\begin{equation}\label{temp_profile}
\frac{T(r)}{T_{mg}}=1.35 \frac{(x/0.045)^{1.9}+0.45}{(x/0.045)^{1.9}+1} \frac{1}{(1+(x/0.6)^2)^{0.45}}
\end{equation}
where $x \equiv r/R_{500}$ is the parametrised radial distance and $T_{mg}$ is the gas-mass-weighted temperature, defined by \cite{2006ApJ...640..691V} as:
\begin{equation}\label{m_t}
R_{500}hh(z)=830 \left( \frac{T_{mg}}{5\text{keV}} \right)^{1.47/3}
\end{equation}
where $h=0.72$ and all other quantities are defined above.  The authors point out that their temperature profiles are self-similar when scaled to the same overdensity radius, which is in agreement with previous authors.  In addition, they cite good agreement between their $M-T$ relation and that produced by other authors \citep[e.g][]{2005A&A...441..893A}.  In this study we will employ the temperature profile described by \cite{2006ApJ...640..691V}.

The theory of cluster magnetic fields suggests that they scale as $B \propto n_e^{1/2}$; further details can be found in Appendix \ref{obs_mag_fields}.


\section{SIMULATION SETUP}

\subsection{Creating a Universal Pressure Profile (UPP) model atmosphere}

\begin{figure}
\begin{center}
\includegraphics[width=0.49\textwidth]{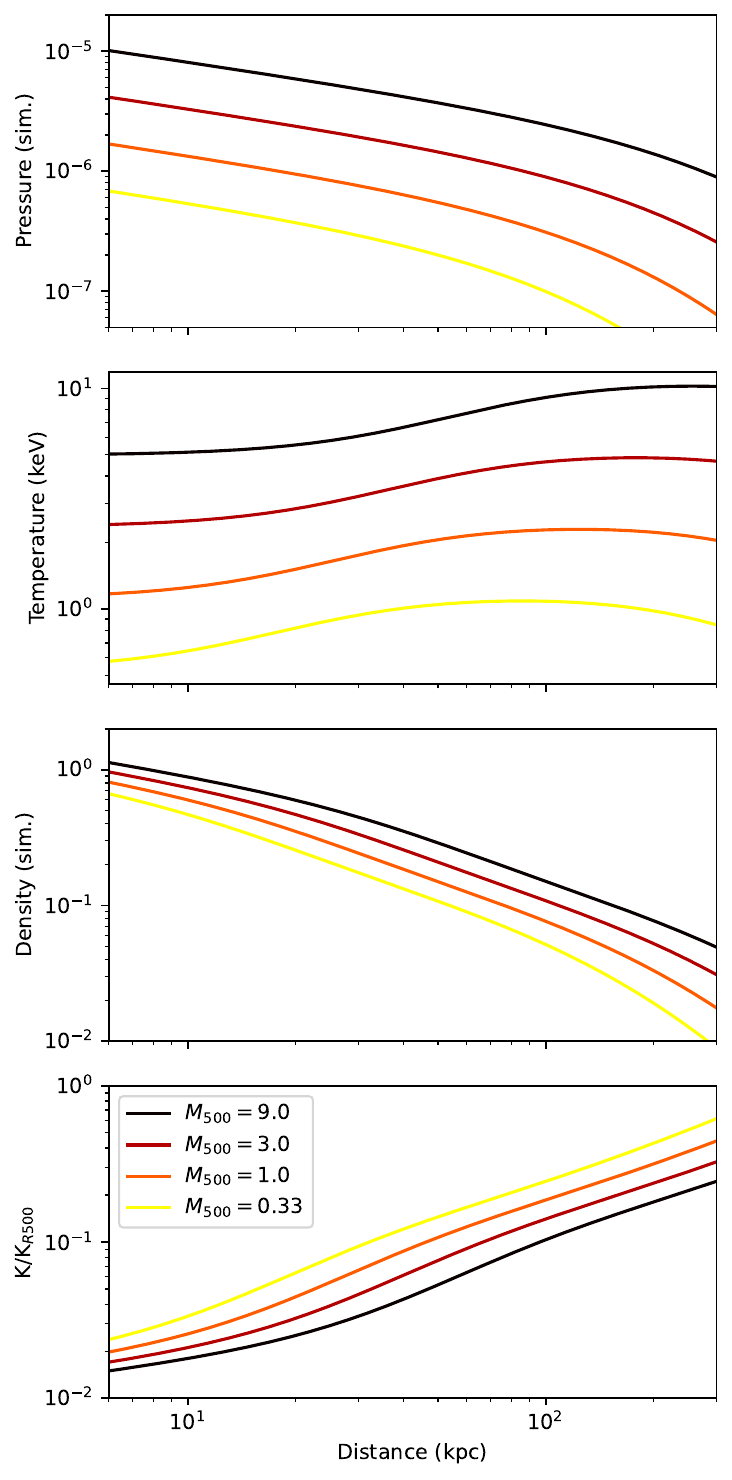}
\end{center}
\caption{UPP pressure (top), temperature, density and entropy (bottom) profiles for the values of $M_{500}$ (in units of $\times \num{e14}\text{h}^{-1}_{70}M_{\odot}$) used in this study.  Distance is from the centre of the cluster.  Pressure and density are measured in simulation units, in which $p_0=\num{2.7e-6}\,$Pa and $\rho_0=\num{3.01e-26}\,$g/cm$^3$.  Specific entropy is calculated as $K=Tn_e^{-2/3}$ where $T$ is the temperature and $n_e$ is the electron number density \protect\citep{2005MNRAS.364..909V}; here we show values normalised to the value at a distance of $R_{500}$.  The gradual decrease in temperature with distance beyond the maximum is not easily seen for this range of distance, a more significant decrease would be visible out to $\sim\,$1 Mpc.}
\label{3UPPtheory}
\end{figure}

The UPP is a self-similar profile with one input variable: cluster mass\footnote{We assume that the model cluster is at a distance of $z=0$.}; it represents observed cluster profiles more faithfully than any other model atmosphere.  The cluster mass is implemented as $M_{500}$ in units of $\text{h}^{-1}_{70}\times \num{e14}M_{\odot}$.  From observations \cite{2013A&A...550A.131P} provide values for $M_{500}$ for clusters in the range $1$ to $\num{15e14}\,\text{h}^{-1}_{70}M_{\odot}$; although other authors have found that the UPP is a good fit outside this range (e.g. \cite{2011ApJ...727L..49S} in their study of groups of galaxies in the range $\num{e13}$ to $\num{1.5e14}\,\text{h}^{-1}_{70}M_{\odot}$); examples of clusters towards the upper limit are very rare.  We will use $M_{500}$ values in the range $0.33$ to $\num{9e14}\,\text{h}^{-1}_{70}M_{\odot}$; the parameters are summarised in Table \ref{UPP_atmosphere_parameters}.  Pressure, temperature, density and entropy distributions for the UPP cluster atmosphere for a range of values of $M_{500}$ are displayed in Figure \ref{3UPPtheory}.  These average pressure profiles can be compared with those of individual clusters from the REXCESS sample, upon which they are based \citep{2010A&A...517A..92A, 2021ApJ...908...91H}; as well as temperature profiles \citep{2006ApJ...640..691V} and density profiles \citep{2008A&A...487..431C}.  Furthermore, the variation of entropy with distance is derived from the temperature and density values (see caption) and compares favourably with profiles derived from observations \citep[e.g.][]{2006ApJ...643..730D, 2017A&A...604A.100G, 2018ApJ...862...39B}.

\begin{table*}
\caption{UPP and equivalent $\beta$-profile atmosphere parameters used to create the pressure, temperature and density distributions for a range of realistic cluster atmospheres.  The $M_{500}$ value represents the mass of the cluster; the other six values are input into {\scriptsize PLUTO}.  $P_{500}$, $R_{500}$, $p_0$ and $r_c$ are all input in simulation units of pressure, length, pressure and length; whereas $T_{mg}$ is in Kelvin. The parameters $p_0$, $r_c$ and $\beta$ define the equivalent $\beta$-profile (see Section \ref{discussion}).  Each run has a label (e.g. jetXX\_haloYY) where XX represents the power of the jet with values of $0.5, 1, 2$ or $\num{4e38}$W; which are indicated by XX having values of $05, 10, 20$ and $40$ (see Table \ref{run_table}) and YY indicates the mass as shown in this table.}
\label{UPP_atmosphere_parameters}

\centering
\begin{tabular}{cccccccc}
\hline
 run & $M_{500}$ & $P_{500}$ & $R_{500}$ & $T_{mg}$ & $p_0$ & $r_c$ & $\beta$\\
 & ($\times\num{e14}\text{h}^{-1}_{70}M_{\odot}$) & (sim.) & (sim.) & (K) & (sim.) & (sim.)\\
 \hline
 jetXX\_halo03 & 0.3333 & $\num{2.2598e-8}$ & 232.90 & $\num{1.0076e7}$ & $\num{5.03891e-7}$ & 16 & 0.48\\
 jetXX\_halo10 & 1 & $\num{4.7007e-8}$ & 335.91 & $\num{2.1276e7}$ & $\num{9.2651e-7}$ & 27 & 0.51\\
 jetXX\_halo30 & 3 & $\num{9.7778e-8}$ & 484.46 & $\num{4.4922e7}$ & $\num{1.9059e-6}$ & 39 & 0.50\\
 jetXX\_halo90 & 9 & $\num{2.0339e-7}$ & 698.71 & $\num{9.4849e7}$ & $\num{3.9827e-6}$ & 72 & 0.70\\
 \hline
\end{tabular}
\end{table*}

Following the methodology of \cite{2011MNRAS.418.1621H}, \citetalias{2014MNRAS.443.1482H}, \citetalias{2016MNRAS.461.2025E} and \citetalias{2019MNRAS.490.5807E}; we implement a cluster magnetic field which is multi-scaled, tangled and has a magnitude related to the cluster density profile.  A summary of the techniques used to create this magnetic field can be found in Appendix \ref{model_magfield}.  The pressure and density profiles are interpolated, along with the three magnetic vector potentials, into the {\scriptsize{PLUTO}} domain at the initialisation step.

\subsection{Ensuring the gravitational stability of the UPP model atmosphere}

Starting with the equation for hydrostatic equilibrium for a spherically symmetric cluster, the pressure exerted by the atmosphere $p(r)$ is assumed to be balanced by the dark matter potential $\Phi(r)$ as
\begin{equation}
\frac{dp(r)}{dr} = -\rho(r) \frac{d\Phi(r)}{dr},
\end{equation}
where $\rho(r)$ is the cluster density profile.  Using the equations of the Universal Pressure Profile (see Section \ref{clusteratmospheres}), we derived the following expression for the dark matter potential:
\begin{equation}
\Phi(r)=\frac{kT}{\mu m_u}\ln\left[x^{\gamma}\left(1+\left(c_{500}x\right)^{\alpha}\right)^{\left(\beta-\gamma\right)}\right],
\end{equation}
this can be compared with dark matter potentials for the $\beta$-profile \citep{2005A&A...431...45K} or the NFW-profile \citep{2008gady.book.....B}.  The RMHD module of {\scriptsize{PLUTO}} requires the input of the gravitational field strength to hold the gas in place, this is derived using $\vec{g}=-\nabla \Phi$ to give:
\begin{equation}
\vec{g}=-\frac{k_B T}{\mu m_u c^2}\left[\frac{\left(\beta-\gamma \right)}{\left(R_{500}/(c_{500}r)\right)^{\alpha}+1}+\gamma\right]\frac{\vec{r}}{r^2}
\end{equation}
where $r=\sqrt{{x}^2+{y}^2+{z}^2}$ and $x, y$ and $z$ are distances in the three Cartesian coordinate directions; $m_u$ is the atomic mass unit; $c$ is the speed of light; $k_B$ is the Boltzmann constant and $T$ is the temperature in Kelvin.  In this model it is assumed that the gas has no self-gravity and that it is held in place entirely by the dark matter potential.  The gas fractions ($f_\mathrm{gas}=M_\mathrm{gas}/M_\mathrm{tot}$) of our model clusters range from $0.053-0.051$ (from low to high-mass) which are at the lower end of the range of observed gas fractions \citep[e.g.][]{2022EPJWC.25700046W}  These low values may be in-part due to the debiased values of the UPP parameters we employ which indicate gas pressure values (and so densities) significantly lower than that predicted by the assumption of hydrostatic equilibrium (see Section \ref{UPP_section}); they may also be due to a possible trend in gas fraction with redshift, whereby lower redshift clusters have a lower gas fraction and we have placed ours at $z=0$ (such a relation is hinted at by \cite{2022EPJWC.25700046W}).

The nature of the UPP is that there is a cusp at the centre of the atmosphere; \cite{2010A&A...517A..92A} highlight that at small radii, this is not realistic, and that a reliable distribution is only expected from tens of kpc from the centre.  The difficulty in terms of coding this aspect of the atmosphere is that the large gradient in pressure towards the centre, when discretised, results in irregularities between the interpolated pressure distribution and the body force implemented within {\scriptsize PLUTO}; these irregularities manifest themselves as acceleration of the material, and so the atmosphere is not balanced in the very centre.  The way to get around this problem is to cut off this unstable cusp at a suitable distance from the centre (we used a distance of 4.2 kpc, determined by the resolution used) and created a zone of constant value in that vicinity (i.e. set the pressure value within a certain radius to be equal to the pressure value at that radius).  The corresponding alteration to the body force also needs to be made to stabilise this area (i.e. set the gravitational field strength to zero).  It should be noted that the injection cylinder injects the jet at a distance of 6.3 kpc from the centre, and so the region cut off does not impact the progression of the jet.  We checked the stability of our atmospheres by running our model, without a jet, for 200 simulation time steps (equivalent to $68.5\,$Myr); in this time a typical RMHD jet would have progressed to the edge of the datacube.  The flat distribution at the very centre of the cluster creates a small discontinuity at its edge which settles over time, otherwise the profile is static.

\subsection{Simulation set up}\label{grid_prep}
For consistency, we use the code units of \citetalias{2016MNRAS.461.2025E}.  The simulation unit for density, length and pressure are set to $\rho_0=\num{3.01e-23}\,$kgm$^{-3}$; $l_0=2.1$\,kpc and $p_0=\rho_0c^2 = \num{2.7e-6}$\,Pa, respectively.  Finally, the simulation unit for the magnetic field is calculated from $B_0=c\sqrt{4\pi\rho_0}$, giving \num{1.84}$\,\mu$T.

The simulations are carried out on a static three-dimensional Cartesian grid centred on the origin and extending to a length of $300$ kpc in each direction.  The central patch is a 4.2 kpc cube in width and is represented by 50 grid points in the y and z-directions and 10 grid points in the x-direction; this provides sufficient resolution for the end of the injection cylinder (see below).  Either side of the central patch is a geometrically stretched grid of 200 cells in the y and z-directions and 300 cells along the x-direction.  The resolution along the y and z-directions ranges from 0.084 kpc at the centre to 6.9 kpc at the grid boundary; along the x-direction the resolution ranges from 0.42 kpc at the centre to 2.1 kpc at the grid boundary.  The cell count is, therefore, $(n_x,n_y,n_z)=(610,450,450)$; all outer boundaries are set to `periodic'.

An injection cylinder is positioned in the centre of the grid, with the two jets running along both directions of the x-axis.  The advantage of this set-up is that AGNs are bipolar outflows and it may be that the out-flowing jet from one side will influence the jet on the other through back-flow of the radio lobes \citep[e.g.][]{2003NewAR..47..573K, 2005A&A...431...45K, 2010MNRAS.405.1303A, 2013MNRAS.430..174H,2014MNRAS.443.1482H, 2014MNRAS.439.2903C}.  We use a jet radius of 0.2 simulation units (0.42 kpc); a comparatively small value compared to \citetalias{2014MNRAS.443.1482H}, \citetalias{2016MNRAS.461.2025E} and \citetalias{2019MNRAS.490.5807E}.  As well as providing a more realistic radius, this choice results in higher density injection for the same jet power and this helps get the jet onto the grid and reduces the clouds of ejecta surrounding the injection region (a problem identified in \citetalias{2016MNRAS.461.2025E}); this is particularly important for our current work as the UPP has a higher density at the core than the equivalent $\beta$-atmosphere.  We keep the injection cylinder half-length the same as our previous work at 3.0 simulation units (6.3 kpc).

\citetalias{2019MNRAS.490.5807E} highlighted the problem of material falling into the side of the injection cylinder: this problem increases with the UPP as a result of higher pressure values around the core.  In order to solve this problem, we created a thin layer around the curved surface of the cylinder (an annular cylinder) of 0.05 simulation units (0.105 kpc) and implemented a zero-gradient boundary condition between it and the surrounding material for the magnetic field, density, tracer and (critically) pressure (similar to the method employed by \cite{2018MNRAS.479.5544M, 2020MNRAS.499..681M}).

We used {\scriptsize{PLUTO}} version 4.4-patch2 for this study \citep{2007ApJS..170..228M}; all of the runs were carried out on the University of Hertfordshire High Performance Computing facility.  Each job was run on 384 Xeon-based cores, taking between one and four weeks each. An output file was written by {\scriptsize{PLUTO}} every 50 simulation time steps (every 0.34 Myr).

We use the special relativistic magnetohydrodynamics (RMHD) physics module, HLLD approximate Riemann solvers and a second order dimensionally unsplit Runge-Kutta time-stepping algorithm, with a Courant-Freidrichs-Lewy number of 0.3.  A divergence cleaning algorithm is used to enforce $\nabla \cdot \textbf{B} = 0$; we calculated the absolute value of the relative divergence error (as defined by \cite{2013MNRAS.432..176P}) and found typical values of $\sim\num{e-3}$.  The model assumes a single-species relativistic perfect fluid (the Synge gas) which is approximated by the Taub-Mathew Equation of State \citep{1948PhRv...74..328T,1971ApJ...165..147M}; for numerical stability reasons, shock flattening was enabled through the use of a more diffusive Riemann solver (HLL) and limiter (MIN-MOD); our simulations are non-radiative for both jet and cluster material.

\subsection{Model jet parameters}
The jet is injected with a constant velocity of $0.994985c$; this corresponds to a Lorentz factor of $\gamma=10$, well within the range observed in jets (see Section \ref{intro}).  The jet power $Q_{\text{\scriptsize{RMHD}}}$ of a jet in SI units, from \cite{2016MNRAS.461.2025E};
\begin{equation}
Q_{\text{\scriptsize{RMHD}}}=\pi r_j^2\nu_j \left[\gamma(\gamma-1)\rho_jc^2+\frac{\Gamma}{\Gamma-1}\gamma^2P_j+\gamma^2\frac{B^2}{2 \mu_0}\right],\label{jetpowerR}
\end{equation}
where $r_j$ is the radius of the jet, $\rho_j$ is the jet density, $B$ the magnetic field strength, $c$ is the speed of light, $\mu_0$ is the permeability of free space, $\nu_j$ is the velocity of the jet (in units of the speed of light), $\gamma$ is the Lorentz factor, $\Gamma$ is the adiabatic index and $P_j$ is the pressure of the jet.

\begin{figure}
\includegraphics[width=0.5\textwidth]{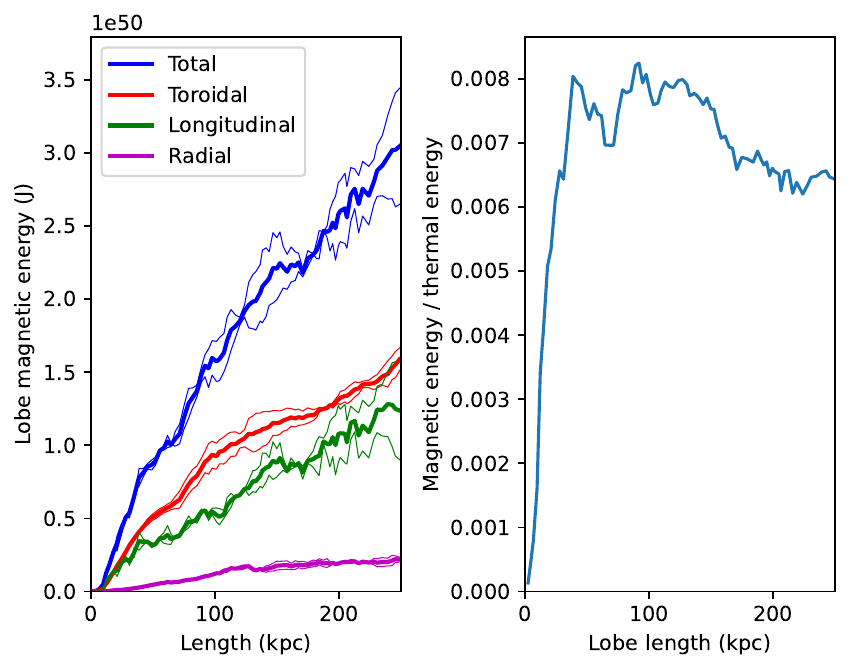}
\caption{Magnetic field evolution in the lobe (LHS) and the ratio between the magnetic and thermal energy in the lobe (RHS) for the fiducial run (jet20\_halo30).  The magnetic field of the jet is injected with both toroidal and longitudinal components (i.e. helical injected field); the values were chosen so that the total magnetic energy of the lobes is $\sim \num{e-2}$ the thermal energy, and the toroidal and longitudinal components are roughly equal in terms of magnetic field energy, once the lobes have reached an average length of $250$ kpc (see text).}
\label{f_test48_magplots}
\end{figure}

As in \citetalias{2016MNRAS.461.2025E} we limit our study to jets with equal contributions of enthalpy and kinetic energy (i.e. the first and second terms of Equation \ref{jetpowerR}).  We inject a helical magnetic field.  The longitudinal component is set to a constant value $B_l$ (for a particular jet power) and the toroidal component is implemented as
\begin{equation}
\text{B}_y=\text{B}_t(\text{z}/\text{r})\qquad \text{and}\qquad \text{B}_z=\text{B}_t(\text{y}/\text{r})
\end{equation}
for $r < r_j$ where $r_j$ is the radius of the jet.  For this study, using jet20\_halo30 as the fiducial run, the jet values of $B_l$ and $B_t$ were determined such that the values of the total magnetic energy of the lobes when they have extended to an average length of $250$ kpc is $\sim \num{e-2}$ the thermal energy (and $B_t$ and $B_l$ contribute equally).  Values for other runs were scaled up and down in proportion to the injected power, all values can be seen in Table \ref{run_table}.  Note that in this model the toroidal component undergoes an unrealistic amplification at early times as the field lines are stretched by the rapidly, outwardly expanding jet.  This is suppressed by gradually increasing the toroidal component from zero to maximum in the first 20 time steps (6.85Myr); from that point the injected field is constant.  The magnetic field evolution for the fiducial run can be seen in Figure \ref{f_test48_magplots}.  The ratio of magnetic energy density to thermal energy density in the lobe for all runs can be seen in Figure \ref{allgraph_magtherm} where it can be seen that this value varies with lobe length, jet power and mass of atmosphere; values range from $\sim \num{2e-3}$ to $\sim \num{3e-2}$ for evolved lobes across the range of parameters studied. 

\begin{table}
\caption{Parameter values for the range of jet powers used in this study; power is measured in watts whereas the jet density ($\rho_j$), pressure ($P_j$), longitudinal ($B_l$) and toroidal ($B_t$) magnetic fields are all measured in simulation units. Each power is run into an atmosphere (marked as YY here) of either $03,10,30$ or $90$ which represent atmospheres of $M_{500} = 0.33,1,3$ and $9\times\num{e14}\,\text{h}^{-1}_{70}M_{\odot}$ (see Table \ref{UPP_atmosphere_parameters}).}
\label{run_table}
\begin{tabular}{ccccccc}
 run & power & $\rho_j$ & $P_j$ & $B_l$ & $B_t$\\
 & $\num{e38}$W & $\num{e-6}$sim. & $\num{e-6}$sim. & $\num{e-3}$sim. & $\num{e-3}$sim. \\
 \hline
 jet05\_haloYY & $0.5$ & 0.6452 & 0.2043 & 0.280 & 1.350\\
 jet10\_haloYY & $1$ & 1.2903 & 0.4091 & 0.403 & 1.909 \\
 jet20\_haloYY & $2$ & 2.5806 & 0.8175 & 0.570 & 2.700 \\
 jet40\_haloYY & $4$ & 5.1612 & 1.6358 & 0.806 & 3.818\\
 \hline
\end{tabular}
\end{table}

\subsection{Postprocessing}
Following \citetalias{2016MNRAS.461.2025E}, the jet is injected with a conserved tracer quantity of value of 1.0 (and zero elsewhere).  Lobes are defined by tracer values $>10^{-3}$.  The bow shock surface (between the shock and the undisturbed ambient medium) is identified in a similar way to the tracer, by line-tracing from the edges of both sides of the volume towards the centre and finding where the radial velocity exceeds the defined value of $75$ kms$^{-1}$.

\section{RESULTS}

\subsection{Dynamics}\label{dynamics}
\begin{figure*}
\includegraphics[width=1.0\textwidth]{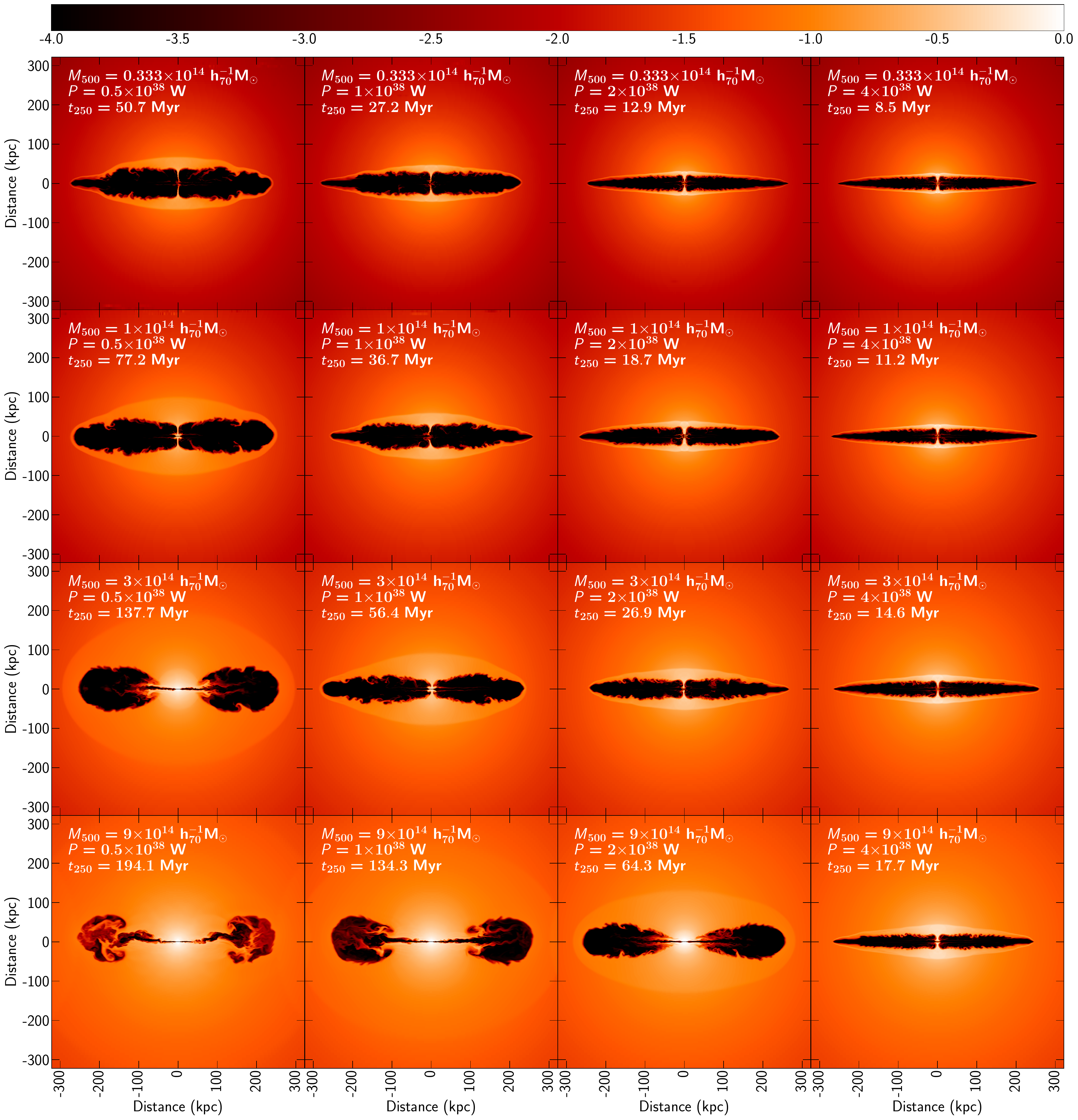}
\caption{Logarithmic density images (xy-slice) for the full range of runs, each presented once the simulated radio galaxy has extended to an average lobe length of 250 kpc.  Cluster mass $M_{500}$, jet power $P$ and time to reach an average lobe length of $250$ kpc, $t_{250}$ are shown on each image.}
\label{16density}
\end{figure*}

Figure \ref{16density} shows density images for all runs when they have progressed to 250 kpc, these are taken as slices through the centre of the xy-plane.  Qualitatively, we observe several features also present in our previous simulations, particularly those of \citetalias{2016MNRAS.461.2025E}.  We observe that some simulations have sufficient buoyancy to lift lobe material clear of the central regions of the simulation, in particular the higher-mass atmospheres and the lower-power jets.  The higher mass atmospheres have greater values of gravitational potential and so will exert a greater force on the rising bubbles - but the lower power jet simulations can clear the central regions of lobe material simply because they take longer to reach 250 kpc and so the buoyancy forces act for longer.  It is clear that the axial ratio varies significantly across these simulations; higher mass atmospheres and lower power jets create lower values for axial ratio by inflating less elongated prolate spheroidal bubbles than higher power and lower mass atmospheres.  Many of our runs have asymmetric lobes, this is due to the slight density perturbations in the initial environment of our models; these are more pronounced than in our previous work as a result of the lower power jets used here (lighter jets are more susceptible to their path being altered by such inhomogeneities).  Lobe asymmetries are a feature of real lobes \citep[e.g.][]{1997MNRAS.288..859H}.

\begin{figure*}
\includegraphics[width=1.0\textwidth]{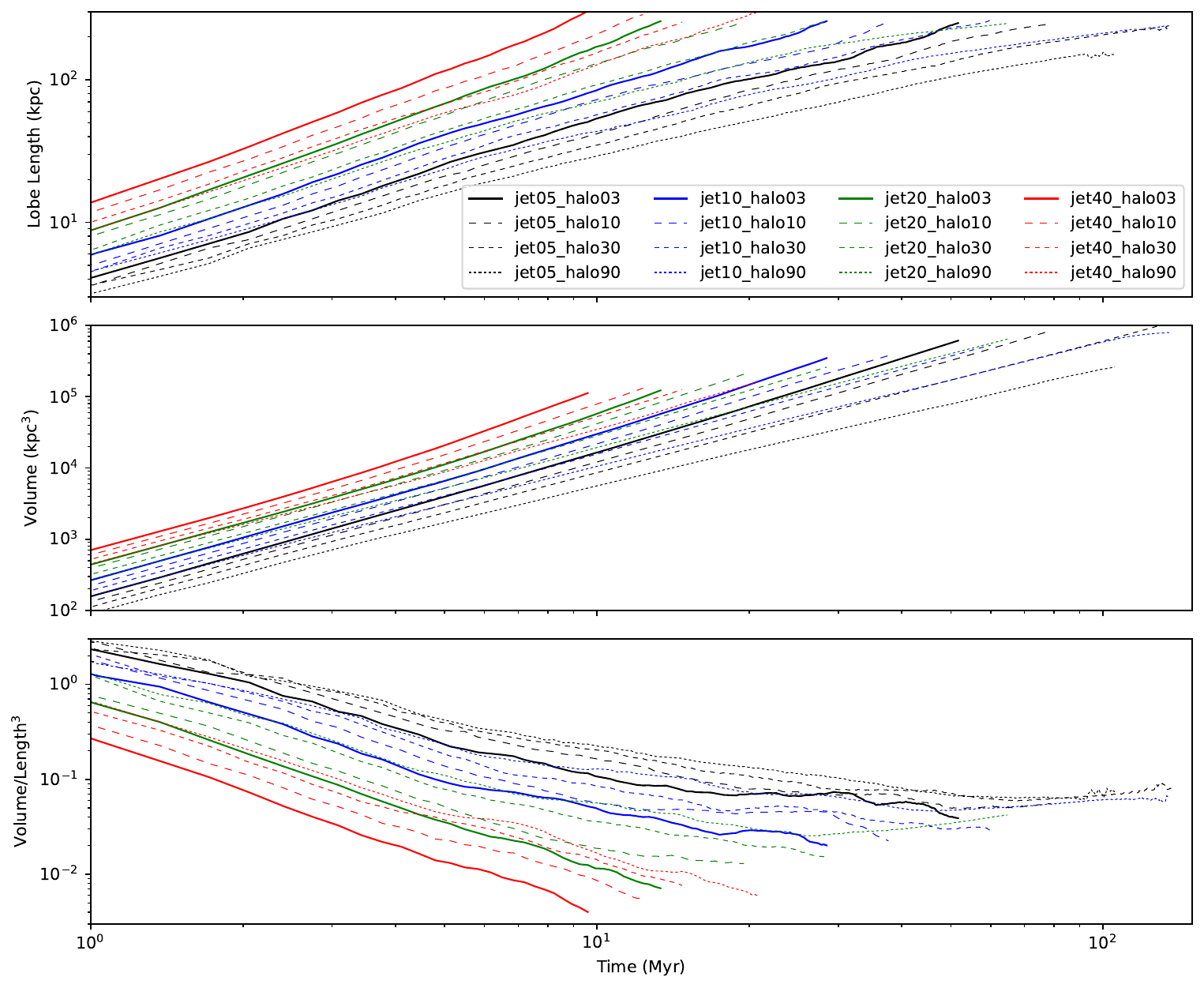}
\caption{Evolution of lobe length (top); volume (middle) and the ratio of volume to the cube of lobe length (bottom) for all runs.}
\label{allgraph_3}
\end{figure*}

Figure \ref{allgraph_3} shows the evolution of the lobe dimensions for all runs in this study and has much in common with both the non-relativistic runs of \citetalias{2014MNRAS.443.1482H} and the mildly-relativistic runs of \citetalias{2016MNRAS.461.2025E}.  The progression of the jet through the cluster is governed by the balance between the lobe pressure and momentum flux of the jet with the density and pressure of the ambient material at the end of the lobe.  As expected, higher power jets progress faster through the same environment as a result of the greater density and so momentum flux of such jets (given we have a fixed injection speed).  Similarly, the richer environments result in the same power jet progressing slower as a consequence of the higher ambient pressure and density of the cluster material. The evolution of lobe length is more linear than our previous work with $\beta$ models; this is because the $\beta$-profile has a near-uniform density at the core which slows the jet until it reaches a steeper density slope further out; whereas the UPP has a decreasing density slope throughout the run as a result of the cusp at the centre.  As expected, the higher power jets accelerate away from the more dense core, but we also found that the higher mass atmospheres (provided the power of the jet was low enough) resulted in the shock front decelerating at longer lengths - corresponding to the inflated lobes seen in the density images of Figure \ref{16density}.

The bottom panel of Figure \ref{allgraph_3} demonstrates that these simulations are not self-similar - a constant value for volume/length$^3$ would be expected for self-similarity.  This finding agrees with our previous work, is a feature of other simulations \citep[e.g.][]{2005A&A...431...45K} and observations \citep[e.g.][]{2000MNRAS.319..562H}.  It is also clearly seen in the density images of Figure \ref{16density} where low power jets moving into rich environments progress more slowly than high power jets moving into poor, and so have lobes of smaller aspect ratio (i.e. are fatter).  We would expect the slower jets to have a smaller aspect ratio as they take longer to reach 250 kpc and so the transverse expansion (which is near-adiabatic) progresses further. It seems likely that KH instabilities also play a role in determining the aspect ratio as the lobes have time to lift away from the cluster core and expose the fast-flowing jet to the near-stationary ambient medium, the jet is no longer protected by the lobe and KH instabilities set in.  These instabilities disrupt the base of the jet and effectively increase the opening angle of the jet-launch, resulting in a larger working surface at the jet termination and so decreases the advance speed of the lobe further.  The impact of the disruption to the jet caused by KH instabilities can be seen in Figure \ref{4mach05}.  

\subsection{Energetics}
\begin{figure}
\includegraphics[width=0.5\textwidth]{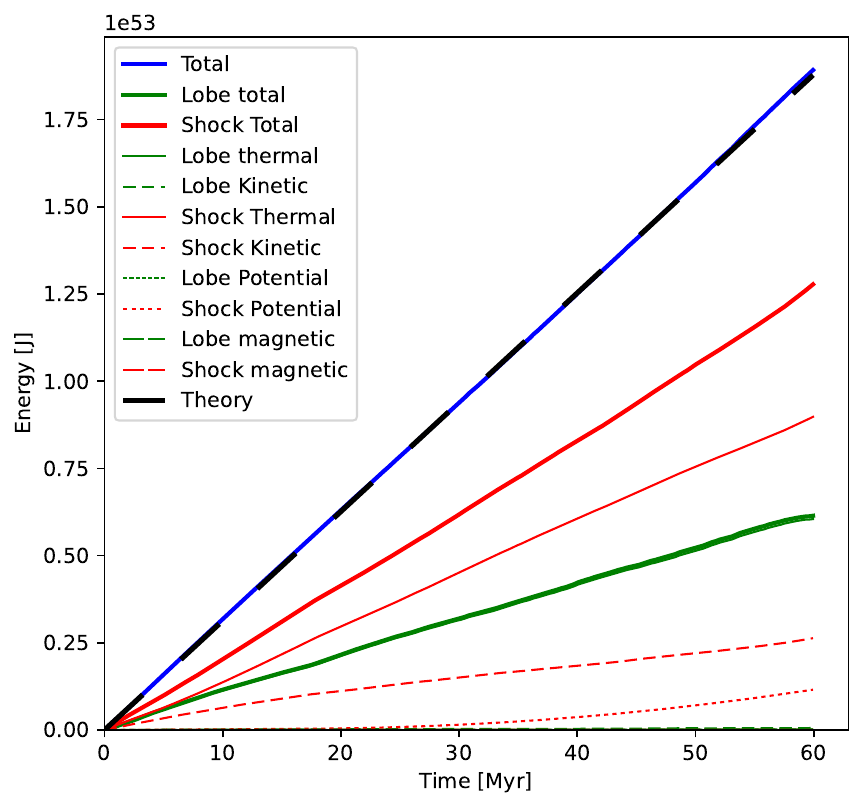}
\caption{Energies in the lobe and shocked regions as a function of time for jet10\_halo30.}
\label{allenergy_f_test54}
\end{figure}

The total thermal, kinetic, magnetic and potential energies contained in the lobe and shocked regions can be plotted against time to investigate the efficiency of the energy injection process.  An example is presented in Figure \ref{allenergy_f_test54} which is for a jet of power = $\num{1e38}$W running into an atmosphere of $M_{500}=3\times\num{e14}\,\text{h}^{-1}_{70}M_{\odot}$.  The total energy measured in the simulation is almost identical to that injected into the system, as expected.

\begin{figure*}
\includegraphics[width=1.0\textwidth]{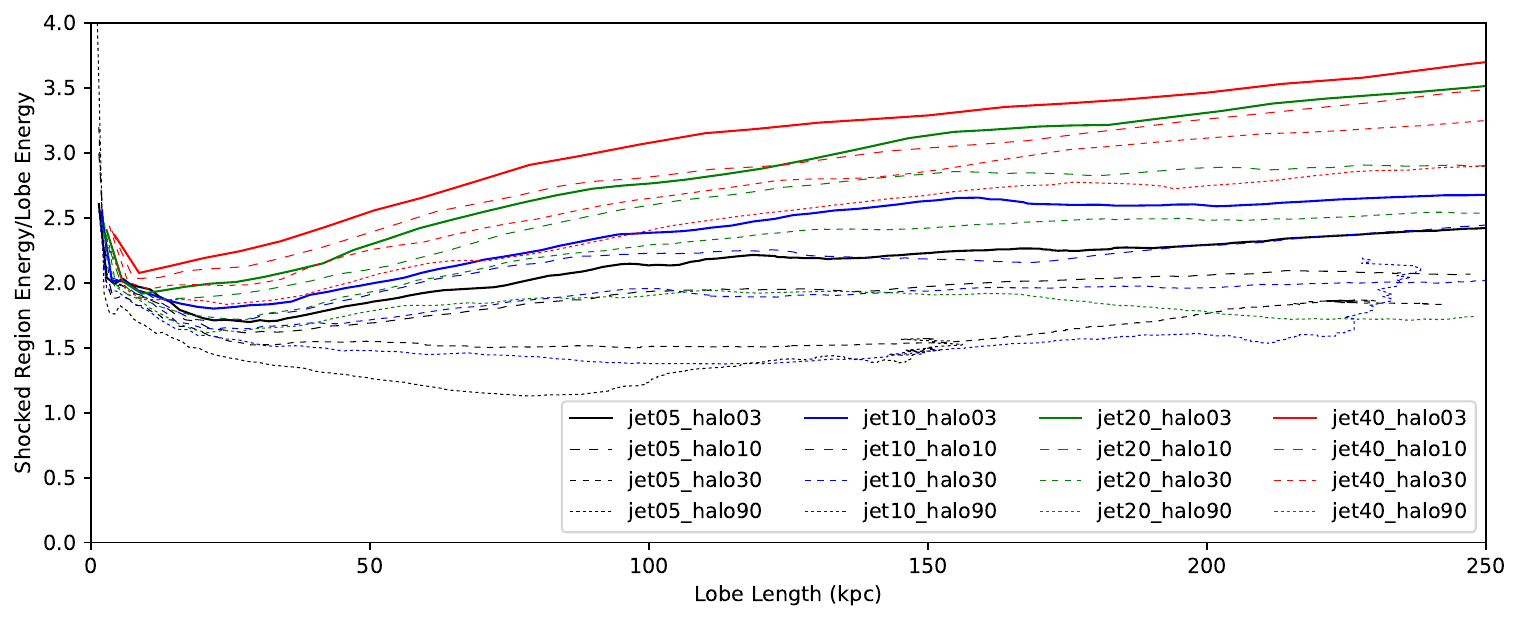}
\caption{The ratio of the energy contained in the shock to that contained in the lobe, against lobe length for all runs.  Low power jets in the highest mass atmospheres (jet05\_halo90 and jet10\_halo90) can be seen to have erratic lines here; this is due to their low shock advance speed such that the method of detection (speed falls below $\num{2.5e-4}c$) is above the advance speed in the outer cluster and so the shock front is not identified accurately; full runs included here for completeness.}
\label{allgraph_shocklobe}
\end{figure*}

The ratio of the energy stored in the shocked region to that stored in the lobes (Figure \ref{allgraph_shocklobe}) shows us that more powerful jets and lower mass atmospheres both result in a greater fraction of the energy going into the shocked region.  The values seen here are generally higher than those of our previous studies; this is likely to be a result of using what are effectively lower mass (and more realistic) atmospheres than those used previously.  The values in this study (ranging from about $1$ to $4$) are double the values we observed in our previous work, although smaller than those found in the simulations by \cite{2011ApJ...743...42P, 2014MNRAS.445.1462P, 2019MNRAS.482.3718P, 2022MNRAS.510.2084P} who find values in excess of $10$; however, their set-up is considerably different to ours and involves a jet perturbation which may influence the transfer of energy from the lobe to the shocked region.  In \citetalias{2013MNRAS.430..174H} and in particular \citetalias{2014MNRAS.443.1482H} we identified that the environment played a role in determining the ratio of shocked to lobe energies: in particular, the greater the density slope the greater the fraction of energy found in the shocked region (i.e. for the $\beta$-profile this was measured by the parameter $\beta$).  For our UPP runs, the density slope is much greater at the core of the cluster and so our current, higher, values can be explained as following this same trend.  As highlighted in Section \ref{dynamics}, KH instabilities impact the lower-power jets moving into richer environments more and result in fatter lobes.  A number of studies suggest that such turbulence in the ICM does not contribute significantly to the heating of the ICM \citep[e.g.][]{2017MNRAS.470.4530W, 2017MNRAS.472.4707B, 2019MNRAS.483.2465M, 2012MNRAS.427.1482G, 2016MNRAS.455.2139H, 2017ApJ...845...91H}.  We conclude that the turbulence in our models does not apper to significantly contribute towards heating of the ICM as our models with the greatest jet disruption result in a smaller fraction of the total energy going into the shocked region.  An alternative explanation is provided by \cite{2017MNRAS.468.3516T}: in their simulations of rapid outbursts they found that more explosive emission events resulted in a greater proportion of the injected energy going into shocks, with up to $\sim 88\%$ for instantaneous/explosive emission and $\sim 0\%$ for the longest and most gentle injection.  The rapidity of the injection, therefore, plays a major role in determining the significance of the heating effect; this agrees with our results as our high-power jets moving into poor environments have the most rapid expansion and the greatest heating effect.

\begin{figure*}
\includegraphics[width=1.0\textwidth]{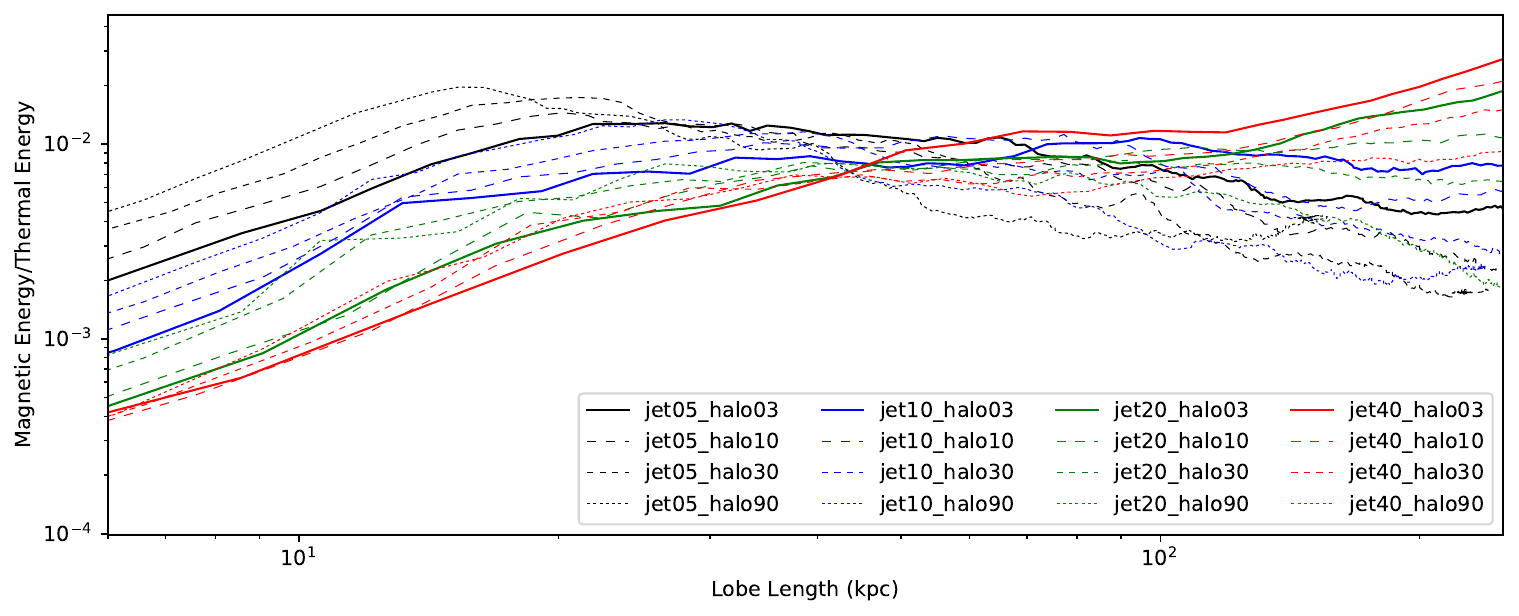}
\caption{The ratio of magnetic energy density to thermal energy density variation with lobe length for all runs.}
\label{allgraph_magtherm}
\end{figure*}

The ratio of magnetic energy density to thermal energy density (Figure \ref{allgraph_magtherm}) varies with power, atmospheric mass and lobe length in a systematic way: all lines appear to be roughly cubic in form; lower power jets have greater values at shorter lobe lengths and then decrease with increasing lobe length, whereas higher power jets increase monotonically.  Atmospheric mass alters the ratio such that for all powers, higher mass increases the ratio at shorter lengths, but decreases it at longer; there is a crossover point on the graph where all atmospheres have the same ratio (roughly) for a particular power.  Through their work in comparing models with observations, \cite{2016A&A...596A..22B} find that the magnetic energy density is two orders of magnitude smaller than the thermal energy with the ratio falling in the range $5-\num{7.5e-3}$; this means that the magnetic fields in clusters are considered not to be dynamically important (i.e. plasma beta $\equiv P_{\text{th}}/P_{\text{B}} \sim 100$).  Figure \ref{allgraph_magtherm} indicates that the magnetic field is not dynamically important, given that all runs have values $\sim 0.01$, although it is possible that it could be locally dynamically important.

\subsection{Synchrotron emission and polarimetry}\label{synchrotron_section}
We simulate Stokes $I$, $Q$ and $U$; and from these we calculate the polarisation and the magnetic field direction.  We follow the method employed in \citetalias{2014MNRAS.443.1482H} and \citetalias{2016MNRAS.461.2025E}; a summary of this can be found in Appendix \ref{stokes_theory}.

\begin{figure*}
\includegraphics[width=1.0\textwidth]{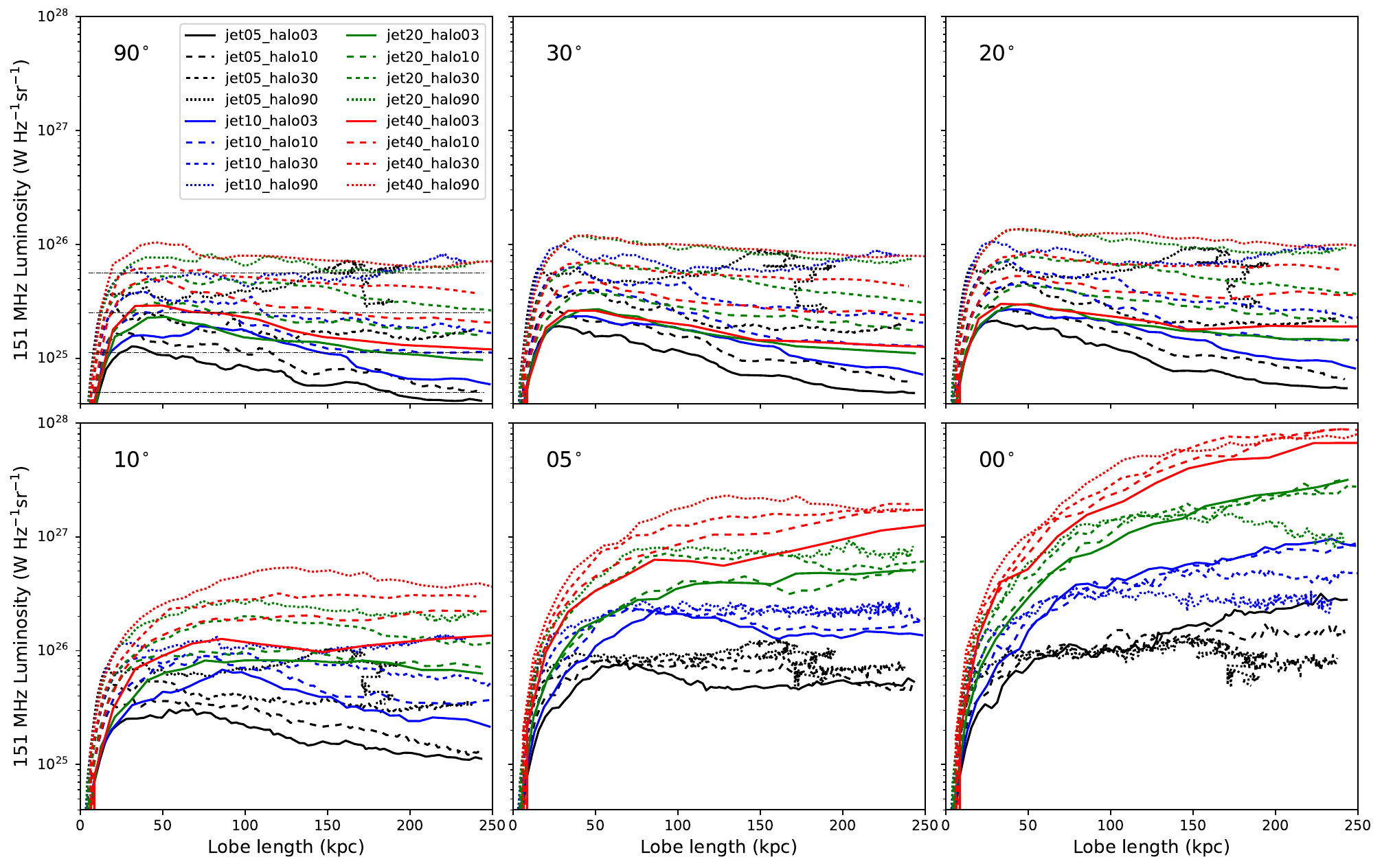}
\caption{Evolution of synchrotron luminosity with lobe length for all runs viewed at various angles to the jet axis as indicated: 90$^\circ$ (top left), 30$^\circ$, 20$^\circ$, 10$^\circ$, 5$^\circ$ and along the jet axis (0$^\circ$).   The black horizontal dashed lines on the top-left plot represent the emission predicted by \protect\cite{1999MNRAS.309.1017W}, which has been normalised to the $\num{1e38}$W lower cluster mass lines at late times (see text).  The line for lowest power run in the highest mass atmosphere (jet05\_halo90) becomes erratic once it has progressed beyond 150 kpc; this jet finds difficulty in punching its way through the ambient medium and the tracer mixes more than other runs and falls below the threshold of $\num{e-3}$, and so the location of the lobe becomes unreliable and is included here for completeness.}
\label{6synch_power}
\end{figure*}

By integrating the synchrotron emission over the whole of the source for each simulation data cube we produce graphs for the total luminosity against lobe length; Figure \ref{6synch_power} shows the results of such an analysis for the full range of power and atmospheric mass used in this study when viewed at various angles to the jet axis.  In line with similar plots from \citetalias{2014MNRAS.443.1482H} and \citetalias{2016MNRAS.461.2025E}, at $90^{\circ}$ the luminosity initially increases with lobe length and then begins to level out, sometimes reaching a maximum or a plateau.  In keeping with \citetalias{2014MNRAS.443.1482H} we find that that atmosphere used impacts the luminosity (in \citetalias{2014MNRAS.443.1482H} the parameters $r_c$ and $\beta$ for the $\beta$-profile were varied).  The general trend is that for higher mass atmospheres the higher the synchrotron output. On our $90^{\circ}$ chart we have included horizontal lines based upon the relation between jet power and the synchrotron luminosity at 151 MHz by \cite{1999MNRAS.309.1017W}; this relation predicts that $Q \propto L_{151}^{6/7}$.  Here we have normalised this to our results for jet power \num{1e38} W (ignoring the highest mass) at late times.  Whilst there is reasonable agreement, our results suggest a relation with a power greater than 6/7. The fact that our results differ to those of \cite{1999MNRAS.309.1017W} is to be expected as they assumed a power-law atmosphere in their model whereas we used the UPP; and their relationship does not capture the time evolution that we observe.  Clearly this relation would also have to be adjusted to incorporate viewing angle if it were to be used on all our plots.  Varying the angle of view has a significant impact on the emission luminosity, particularly for angles $\sim 1/\gamma$ and less (where $\gamma$ is the Lorentz factor) as Doppler boosting of the jet becomes significant (about $6^{\circ}$ in our study) and the emission from the jet dominates over that of the lobe.  We must remind ourselves that we have made a number of assumptions about the pressure, magnetic field structure and electron energy spectrum of a real jet, and so the results we present here for viewing the jet at small angles must be regarded with caution; nevertheless, end-on jets are observed (e.g. blazars, \cite{2019ARA&A..57..467B}) although we feel that the amount of Doppler boosting is exaggerated in our models.  Movies showing the simulated synchrotron emission at various angles and lobe lengths can be found at \url{http://uhhpc.herts.ac.uk/~ms/synchrotron.html}.

\begin{figure}
\includegraphics[width=0.4\textwidth]{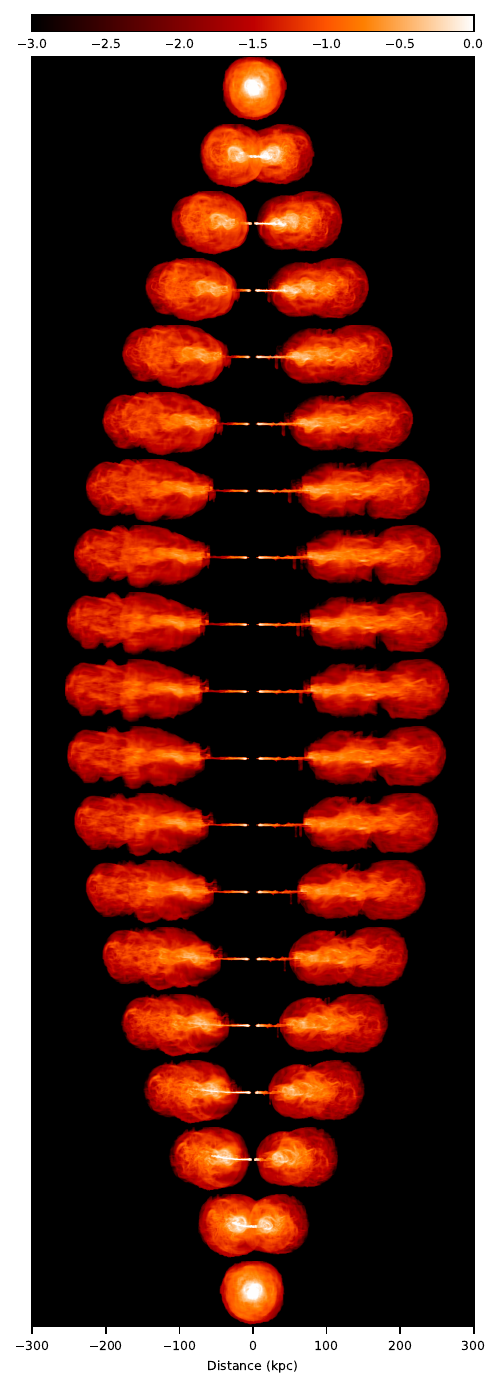}
\caption{Sequence of synchrotron emission maps; top image observing down the jet axis, each subsequent image is positioned an additional $10^{\circ}$ from the axis, the bottom image is looking down the counterjet. Images for a jet of power $\num{2e38}$W running into an atmosphere of $M_{500}=9\times\num{e14}\,\text{h}^{-1}_{70}M_{\odot}$ with jet extended to an average length of 250 kpc; arbitrary logarithmic brightness scale.  Doppler boosting of the jet and counterjet is visible at small angles to the jet-counterjet axis; note that the logarithmic scale makes this appear less pronounced; the true magnitude of the Doppler boosting of the jet is more easily seen in Figure \ref{6synch_power}.}
\label{synch180}
\end{figure}

\begin{figure*}
\includegraphics[width=1.0\textwidth]{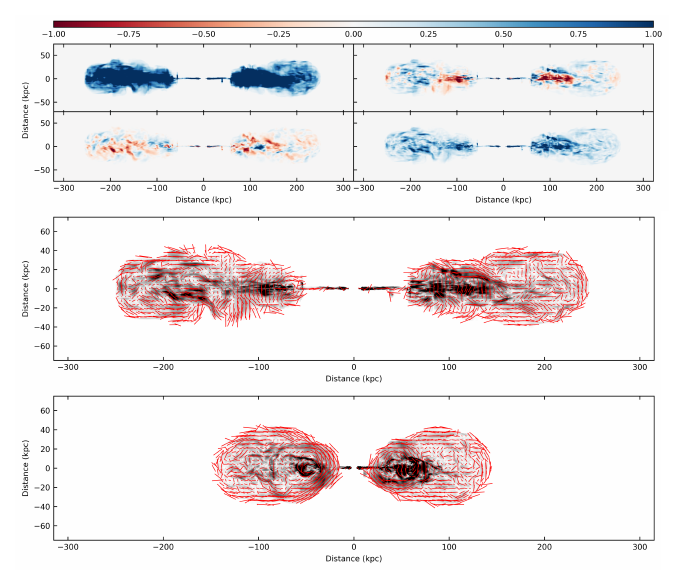}
\caption{Synchrotron emission maps for jet20\_halo90 with lobe extended to 250 kpc.  Top panel: (from top left) Stokes $I$ (synchrotron emission), $Q$, $U$ and the polarised intensity $P=\sqrt{Q^2+U^2}$ (bottom right).  Middle and Lower panels show  synchrotron emission maps overlaid with vectors representing the direction of the magnetic field and the magnitude of the linear polarisation.  Middle panel viewed at $90^{\circ}$ to the jet axis and Lower panel at $30^{\circ}$.  Arbitrary brightness scale used.}
\label{stokes_images}
\end{figure*}

We created simulated synchrotron two-dimensional emission maps; an example can be seen in Figure \ref{synch180} where the angle of view has been progressively increased in order to observe the simulated Doppler boosting.  We also created emission maps for Stokes $Q$ and $U$ and $P$ (polarised intensity); examples can be seen in the top panel of Figure \ref{stokes_images} and are very similar to those produced by our previous work in \citetalias{2014MNRAS.443.1482H} and \citetalias{2016MNRAS.461.2025E}, although a greater level of fine detail is revealed as a consequence of the higher resolution used in the current study.  We can overlay a synchrotron emission map with vectors with magnitude that represents the magnitude of the fractional polarisation and with a direction corresponding to that of the magnetic field.  Such an image viewed at $90^{\circ}$ to the jet axis is presented in the middle panel of Figure \ref{stokes_images} and shows the same trends found in to our previous work in that the evolved lobe polarisation is higher at the edges where the magnetic field direction is parallel to the lobe-shock boundary.  Furthermore, in the bottom panel of Figure \ref{stokes_images} we demonstrate that these findings are also true when observing from an angle (here at $30^{\circ}$ to the jet axis), in common with what was found in the simulations of \cite{2011MNRAS.418.1621H} and seen in observations by \cite{1997MNRAS.288..859H}.  These similarities give us confidence that the conclusions drawn from our previous work remain valid for these more relativistic and higher resolution studies which use a more realistic helical magnetic jet injection.

\section{DISCUSSION}\label{discussion}
\subsection{Comparisons with the {\boldmath$\beta$}-model}
Here we draw comparisons between our UPP models and $\beta$-models in order to establish whether our UPP model leads to different results from the established $\beta$-profile.  We also compare our results with observations of Doppler boosting and observations of radio jets, and report on our resolution study.

There is no established way to create an equivalent $\beta$-profile from a UPP profile: \cite{2018MNRAS.475.2768H} present a graph comparing these two profiles, and it is clear that they differ considerably at the cluster centre in particular.  Starting with a UPP profile, we derived the equivalent $\beta$-profile using an iterative technique beginning with the general profile of $\rho_0=1.0$ simulation units, $r_c=0.1R_{500}$ and $\beta=0.5$; and then varied each of these three parameters by a small amount and minimised the difference between the two profiles.  The UPP profile has a `core radius', defined by \cite{2010A&A...517A..92A} as $r = 0.1R_{500}$; there is no indication that this is equivalent to the $\beta$-profile $r_c$ but it gives an order of magnitude value to start with; the initial $\rho_0$ and $\beta$ are typical values employed by  \citetalias{2014MNRAS.443.1482H} and \citetalias{2016MNRAS.461.2025E}.  The `best fit' $\beta$-model values we derived for each UPP model atmosphere can be seen in Table \ref{UPP_atmosphere_parameters}.

\begin{figure}
\includegraphics[width=0.49\textwidth]{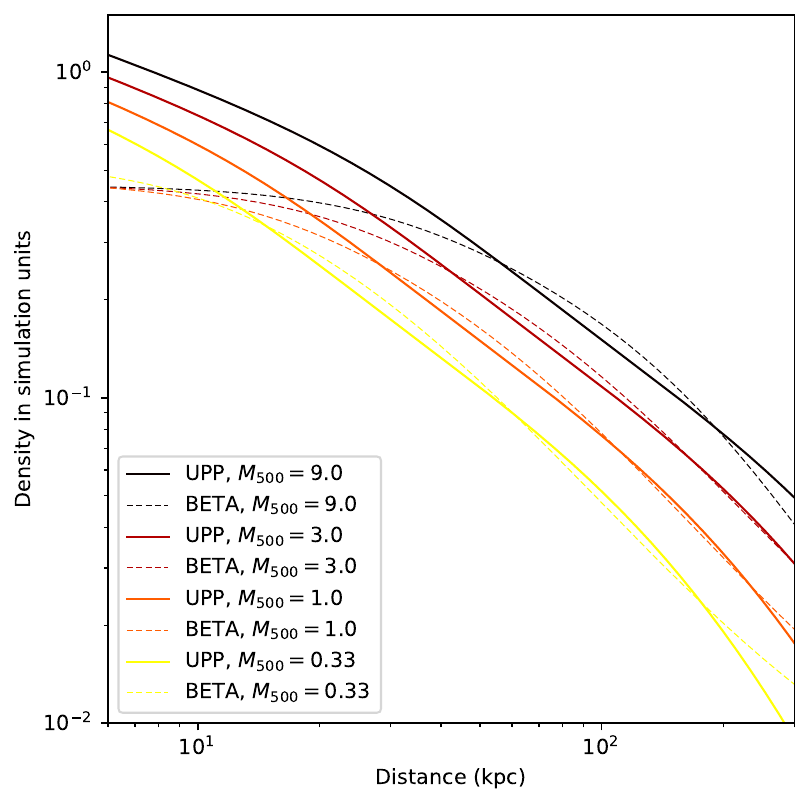}
\caption{Fitting the $\beta$-profile to the UPP by minimising the difference between the two curves (by varying the $\beta$-model parameters $\rho_0$, $r_c$ and $\beta$).  Continuous lines are for the UPP atmospheres and dashed lines are for the corresponding $\beta$-profile.  Note that the log-log plot used here highlights the difference at the core.}
\label{BETAUPPgraph}
\end{figure}

Density profiles of the UPP and corresponding $\beta$-model for the range of cluster mass used in this study can be seen in Figure \ref{BETAUPPgraph}.  The two models differ most at the cluster centre and follow very similar curves out to the limit of this study (i.e. 300 kpc). Beyond this range of study, the cluster density is comparatively low for both models and so any variation at large distances would have an insignificant impact on the progression of the jet and lobe in comparison with moving through the higher density regions closer to the core.

\cite{2010A&A...522A.105G} provide $\beta$-model parameters for the much-studied Coma cluster of $\rho_0=\num{3.5e-3}\,$m$^{-3}$, $r_c=245\,$kpc and $\beta=0.654$.  We would not expect these to be exactly replicated in an `average model' such as the UPP, although one would hope that values would be the same order of magnitude.  Assuming a Coma cluster mass with $M_{500}=\num{2.9e14}\text{h}^{-1}_{70}M_{\odot}$ \citep{2019MNRAS.485.2922L}, and so using our cluster mass of $M_{500}=\num{3e14}\text{h}^{-1}_{70}M_{\odot}$, as well as the general temperature profile used in this study (see Section \ref{temperature_variation}) which has a central temperature value of 2.4 keV, we read off the parameters from Table \ref{UPP_atmosphere_parameters} and converting from simulation units we obtain: $\rho_0=\num{1.4e-3}\,$m$^{-3}$, $r_c=95\,$kpc and $\beta=0.54$ and so we do have the correct order of magnitude and values derived from the UPP which differ from observed values by no more than a factor of $\sim\!3$.

\begin{figure}
\includegraphics[width=0.5\textwidth]{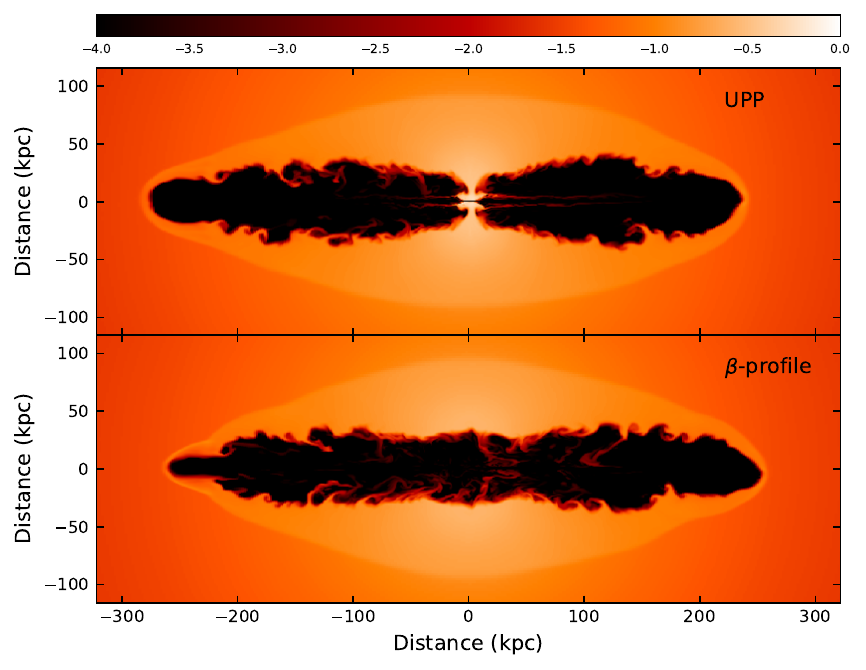}
\caption{Logarithmic density images (xy-slice) for a jet of power $\num{1e38}$W and moving into a UPP atmosphere of $M_{500} = 3\times\num{e14}\,\text{h}^{-1}_{70}M_{\odot}$ (top) and the same power jet moving in to the equivalent $\beta$-atmosphere (bottom).  Parameters used to create these two atmospheres can be seen in Table \ref{UPP_atmosphere_parameters}.  Both models shown have progressed to 250 kpc and take almost exactly the same time to do this (56 Myr for jet10\_halo30 and 57 Myr for beta10\_halo30).}
\label{2images_prod_beta}
\end{figure}

\begin{figure}
\includegraphics[width=0.48\textwidth]{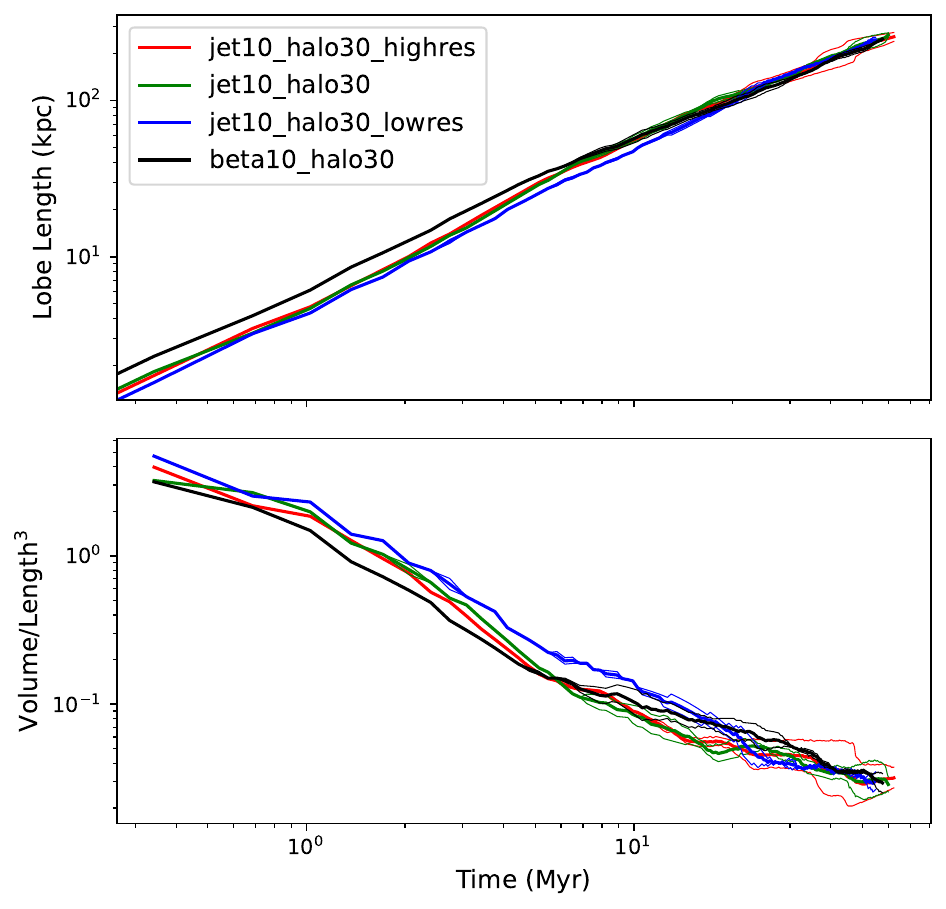}
\caption{Evolution of length (top) and the ratio of volume to length$^3$ (bottom) for jet10\_halo30 (UPP run), beta10\_halo30 ($\beta$ run); jet10\_halo30\_highres (UPP high resolution run) and jet10\_halo30\_lowres (UPP low resolution run).  See Figure \ref{2images_prod_beta} for density images of jet10\_halo30 and beta10\_halo30.  The thick line is the average for the two lobes and the thin lines are for individual lobes; this gives a sense of the scatter between the lobes.}
\label{2graph_comp1}
\end{figure}

\begin{figure}
\includegraphics[width=0.48\textwidth]{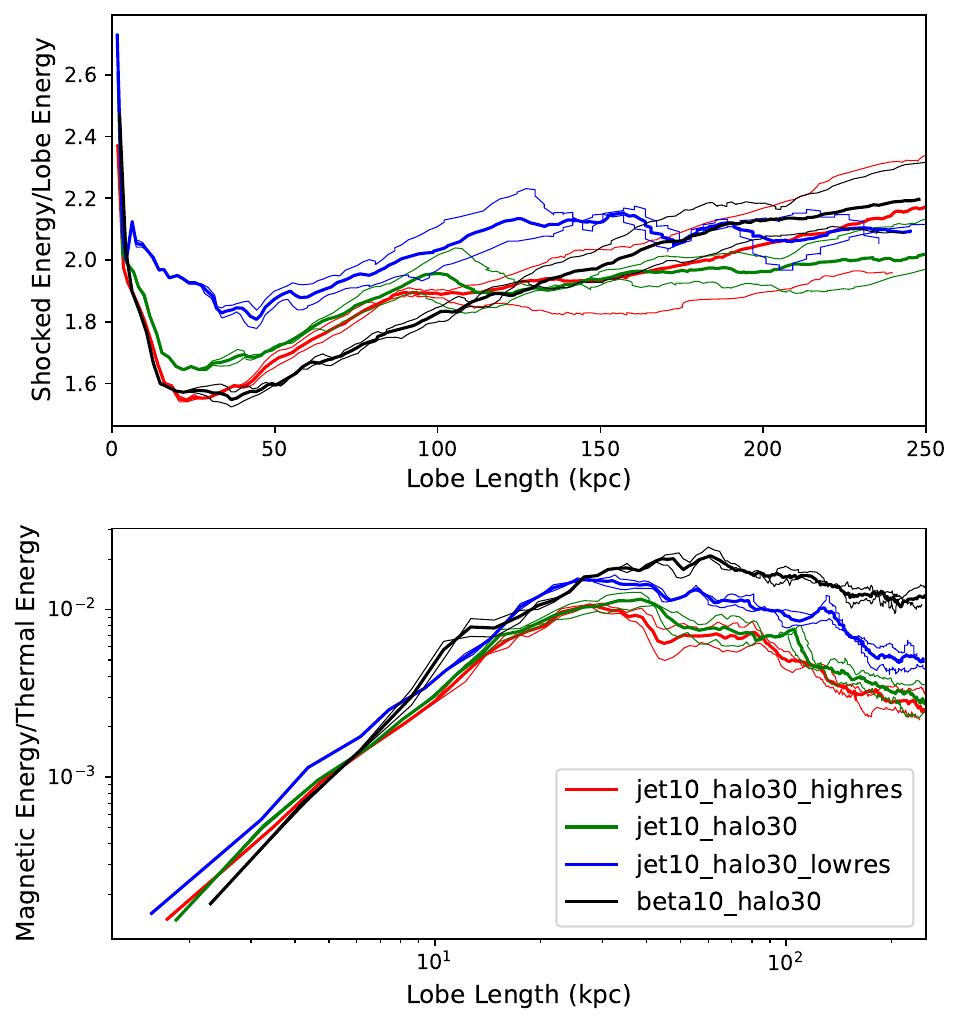}
\caption{Evolution of the ratio of shocked region to lobe region energy (top) and lobe magnetic to thermal energy (bottom) for jet10\_halo30 (UPP run), beta10\_halo30 ($\beta$ run); jet10\_halo30\_highres (UPP high resolution run) and jet10\_halo30\_lowres (UPP low resolution run).  See Figure \ref{2images_prod_beta} for density images of jet10\_halo30 and beta10\_halo30.  The thick line is the average for the two lobes and the thin lines are for individual lobes; this gives a sense of the scatter between the lobes.}
\label{2graph_comp2}
\end{figure}

We ran a jet of power $\num{1e38}$W into a $M_{500} = 3\times\num{e14}\,\text{h}^{-1}_{70}M_{\odot}$ atmosphere (jet10\_halo30) and the same power jet into the equivalent $\beta$-atmosphere (beta10\_30) (see Table \ref{UPP_atmosphere_parameters} for parameter values).  Both runs extended to 250 kpc and density images can be seen in Figure \ref{2images_prod_beta} with dynamic and energetic comparisons presented in Figures \ref{2graph_comp1} and \ref{2graph_comp2}.  The atmosphere model chosen impacts the morphology of the lobe as well as its dynamics and energetics.  The UPP is more able to clear lobe material from the central regions and so leads to a more distinct pair of bubbles whereas the $\beta$-model forms one elongated shape where the lobes remain merged at the centre.  However, given the rich variety of observed lobe shapes, examples can be found which correspond to both shapes seen here: 3C 438 has lobes which are more separated whereas 3C 20 has lobes which merge near the core (radio images of these lobes and others can be seen in \cite{1997MNRAS.288..859H}).  The $\beta$-model has a higher aspect ratio at early times (Figure \ref{2graph_comp1}) although differences become negligible once lobe lengths of hundreds of kpc are reached; this is consistent with the atmosphere models given that the major difference between $\beta$ and UPP is at the core.  Comparisons are more marked for the evolved lobes when considering the energetics: the top chart of Figure \ref{2graph_comp2} shows that the lobe in the $\beta$-model passes on energy to the shocked region less efficiently at early times but ends up transferring the greatest by 250 kpc.  Similarly the magnetic/thermal ratio of the lobe is lower at early times but ends up being greater at late times (Figure \ref{2graph_comp2} (bottom)).  The values for individual lobes have been plotted on Figure \ref{2graph_comp2} (and \ref{2graph_comp1}) and these give an indication of the variability within these runs; this demonstrates that there is a genuine difference between the UPP and $\beta$-model atmospheres which impacts their energetics, particularly at late times.

\subsection{Synchrotron imaging}
\begin{figure}
\includegraphics[width=0.5\textwidth]{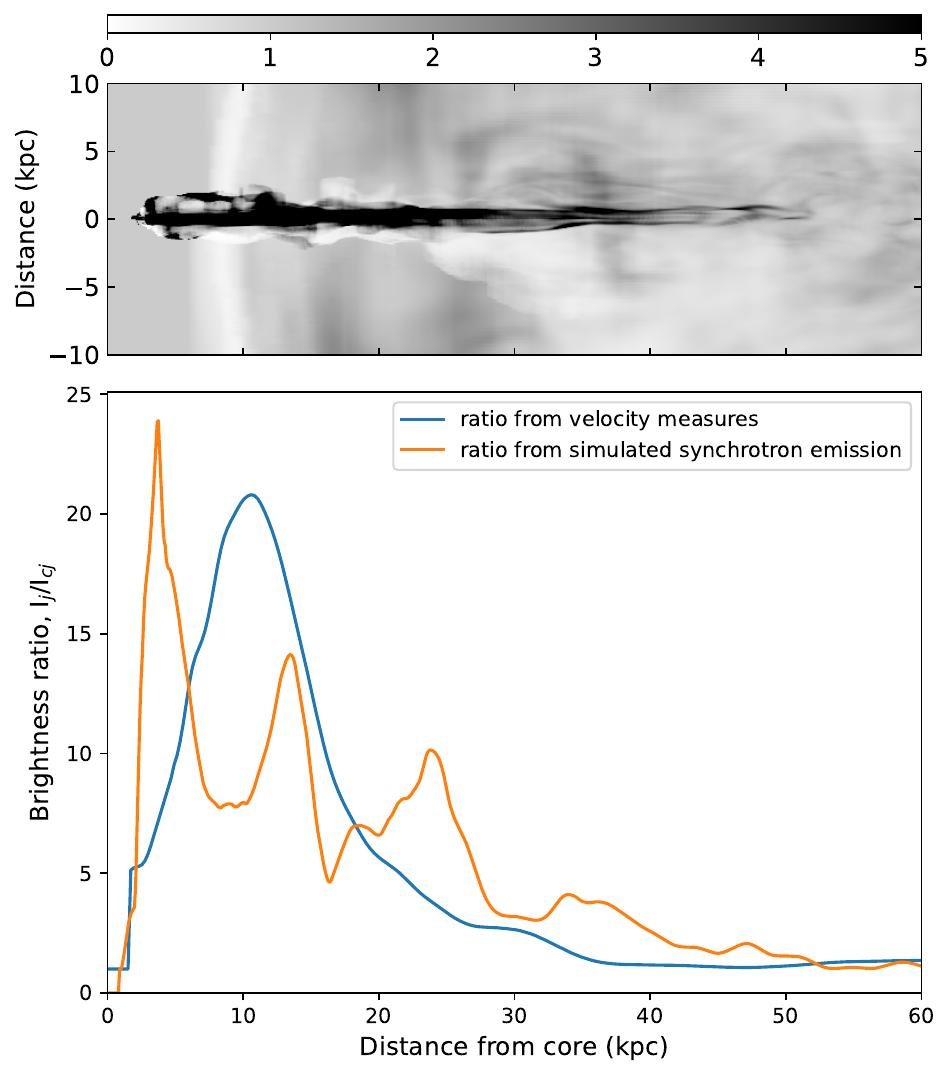}
\caption{A grey-scale image of jet sidedness (top) for jet05\_halo30 viewed at $30^{\circ}$ to the jet axis.  It is constructed by rotating the simulated synchrotron image by $180^{\circ}$ about the nucleus and dividing by itself.  Values above 5 are saturated in this image.  Longitudinal jet sidedness profiles (bottom) derived from the simulated image of jet sidedness (shown above) and also from the simulated bulk velocity measures (see text).}
\label{brightnessratio}
\end{figure}

As seen in Section \ref{synchrotron_section}, our model reproduces the expected Doppler boosting of the jet when viewed along (or at small angles to) the jet axis.  This effect can be measured by the brightness ratio ($I_j/I_{cj}$) and a plot of jet base sidedness can be created from observations by rotating the synchrotron image of a bipolar jet by $180^{\circ}$ about the nucleus and dividing the jet image by the counterjet \citep[e.g.][]{1996IAUS..175..147L, 1997MNRAS.288L...1H}.  We copy this methodology for the simulated emission of jet05\_30 viewed at $30^{\circ}$ to the jet axis to obtain Figure \ref{brightnessratio} (top).  This ratio of jet base sidedness can also be derived from the jet bulk velocity $\beta$ and viewing angle $\theta$ as \citep[e.g.][]{1996IAUS..175..147L}
\begin{equation}\label{sidedness}
\frac{I_j}{I_{cj}}=\left[\frac{1+\beta \cos\theta}{1-\beta \cos \theta} \right]^{2+\alpha}
\end{equation}
where $\alpha$ is the synchrotron spectral index (taken to be 0.5).  This formula assumes isotropic emission in the rest frame and so can only be used as an approximation.  From our simulation we calculate the brightness ratio using both the left hand (simulated synchrotron brightness ratio) and right hand (bulk velocity) sides of Equation \ref{sidedness} to produce Figure \ref{brightnessratio} (bottom).  The line derived from the bulk velocity falls from the initial maximum more smoothly than that derived from the brightness ratio, this is to be expected as an isotropic magnetic field has been assumed for the bulk velocity calculation, whereas the brightness ratio calculation incorporates the variation in the simulated magnetic field.  A very similar evolution for the simulated brightness ratio was found by \cite{1996IAUS..175..147L} (their Figure 4) from their synchrotron observations of 3C 31, a decelerating jet.  Measurements of the bulk velocity of our jet show that it is also decelerating in this inner region; this is as a result of buoyancy having lifted the lobe and exposed the jet to the ambient material which is entrained into the edge of the jet and so slowing it down.

\begin{figure}
\includegraphics[width=0.5\textwidth]{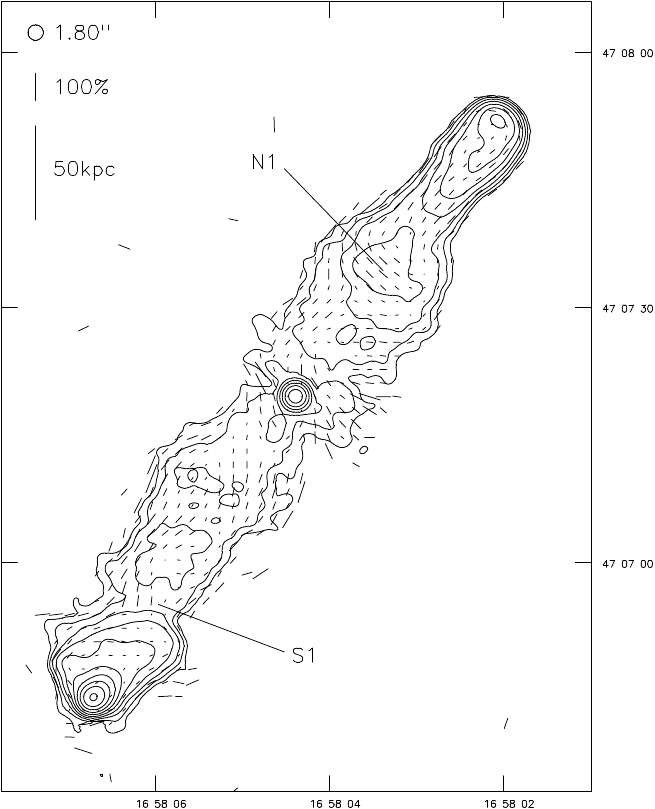}
\caption{3C349 at 1.80-arcsec resolution.  Contours at 0.25$\times$(-2,-1,1,2,4,....) mJy beam$^{-1}$.  Image taken from {\protect\cite{1997MNRAS.288..859H}}.}
\label{3C349}
\end{figure}

Figure \ref{3C349} shows an image of the radio galaxy 3C 349 (reproduced from \cite{1997MNRAS.288..859H}); this can be compared with the synchrotron image of Figure \ref{stokes_images} (top left), the images of Figure \ref{synch180} or the density images of Figure \ref{16density}.  Favourable similarities can be noted including: the approximate axial ratio, the effect of buoyancy which thins out the radio-emitting lobe from the central regions, the rough large-scale symmetry and the overall length.  Note that the example chosen (3C 349) is imaged close to $90^{\circ}$ to the jet axis as otherwise Doppler boosting would complicate comparisons with the density images presented here.  The presence of Kelvin-Helmholtz instabilities on the lobe surface and the small-scale morphological variability is evident in our simulated density and synchrotron images as a result of density inhomogeneities in the ambient cluster; these are also evident in the observed radio image.  In particular, some of our lower power runs produce fatter lobes in the vicinity of the cluster centre, but then the lobe growth increases once it reaches lower density material in the outer cluster and subsequently produces a `neck', a similar feature is seen on the upper lobe of 3C 349.

\subsection{Lack of hotspots}
\begin{figure*}
\includegraphics[width=1.0\textwidth]{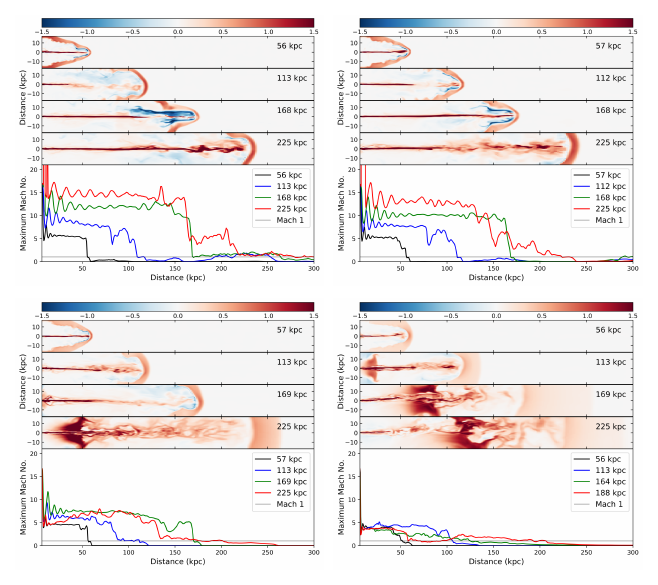}
\caption{Internal Mach number ($\Gamma \beta/\Gamma_s \beta_s$) for a jet of power $\num{0.5e38}$ W running into an atmosphere of $M_{500}$ (measured in units of $\times\num{e14}\,\text{h}^{-1}_{70}M_{\odot}$): $0.3$ (top left), $1$ (top right), $3$ (bottom left) and $9$ (bottom right). Images are of x-y slices through the midpoint and the plots are of the maximum Mach number in the y-z plane at that distance from the core (to take into account the movement of the jet out of the x-y slice shown).}
\label{4mach05}
\end{figure*}

As discussed in our earlier work (\citetalias{2016MNRAS.461.2025E}), our models do not reproduce the hot spots characteristic of `edge brightened' FR II radio galaxies (these are clearly visible in observations such as Figure \ref{3C349}).  Knots and hotspots are synchrotron emission created by highly energetic particles; these features are normally far from the injection and recollimation sites near the central nucleus such that adiabatic losses \citep{1973MNRAS.164..243L} and spectral ageing \citep[e.g.][]{1962SvA.....6..317K} would prevent particles of sufficient energy reaching them so that there needs to be local acceleration of particles, such as the process of diffusive shock acceleration (DSA, \cite{1978MNRAS.182..147B}).  We mapped the internal Mach number of our simulations: $\Gamma \beta/\Gamma_s \beta_s$; where $\Gamma$ is the Lorentz factor, $\beta$ is the bulk velocity in natural units and the subscript $s$ denotes values for the sound speed (see Figure \ref{4mach05}).  We found that the jet has a Mach number in excess of 1 for almost all its length, and so a shock could take place along the jet right up to the tip of the lobe.  Simulations by \cite{1992A&A...260..243M, 1996ApL&C..34..295M, 1997MmSAI..68..163M} indicate that the evolution of Kelvin-Helmholtz (KH) instabilities at the surface of the jet results in the formation of shocks which may be the cause of knots.  Such disruption of the jet through KH instabilities may also be a factor influencing the termination of the jet in wide-angle tail radio sources \citep{2004MNRAS.349..560H, 2005MNRAS.359.1007H}, particularly where a hotspot is seen at the base of the plume.  In our simulations we can see that shortly before the end of the lobe the Mach number falls to subsonic levels as the jet encounters decelerating material back-flowing into the lobes; and the jet terminus moves around in these snapshot images, in keeping with the description of \cite{1974MNRAS.166..513S} and his `dentist drill' model; we see that this feature is increased when a jet of the same power is injected into a richer environment as this reduces the Mach number and so leads to a less stable jet.  We note also that in the richer environments the lobe lifts away from the core of the AGN through buoyancy and exposes the outer surface of the jet to the higher-density ambient material; this should reduce the stability of the jet and in our models we see significant disruption through entrainment and deceleration of the outer surface of the jet leading to a `sheath' of slower-moving material around the `spine' of the jet; the slower moving sheath is more susceptible to KH instabilities.  We suspect that the creation of this sheath of material may be due to our boundary conditions when launching the jet, resulting in an amplification of the KH instabilities beyond what would be realistic.

\subsection{Aspect ratio}
The aspect ratio of our models varies significantly (see Figure \ref{16density}).  FR II's tend to form prolate spheroidal-shaped lobes with an axial ratio between $1.3$ and $6$ \citep{1984MNRAS.210..929L} with the largest fraction between 1.5 and 2 \citep{2008MNRAS.390..595M}, where axial ratio is defined as the ratio of length to width.  Our lower power jets moving into richer environments fall well within these ranges, although our higher-power jets are at and beyond the upper range.  The jet power and cluster masses of our models match well those seen in observations, and so there must be another feature of high power jets which is missing from our models.  Numerical simulations limit the resolution by means of a minimum grid size and so reduce the growth of instabilities experienced by the model jet as it moves through the material within the lobe which would otherwise slow the jet's progress \citep[e.g.][]{2010MNRAS.402....7M}; this effect explains why the lobe growth is often slowed when higher resolution simulations are conducted in convergence testing \citep[e.g.][]{2001A&A...380..789K, 2019MNRAS.482.3718P}.  Sometimes a method is employed to compensate for the reduced instability growth by limiting the flow of the jet so the model has time to form realistic-shaped lower aspect-ratio lobes.  One such technique is to introduce a helical perturbation to the injection-velocity; this is achieved through applying small velocities normal to the jet flow at the injection cylinder, for example in the simulations of \cite{2019MNRAS.482.3718P, 2022MNRAS.510.2084P} where a sum of normal sinusoidal velocity perturbations of different angular frequencies is used.  They observe a very stable jet (phase I) up to 100 kpc, and beyond this (phase II) a `dentist drill' phase whereby the jet is appreciably disrupted; they refer to this second phase happening in their model as a result of the `wobble' introduced and cite irregularities near the jet head as evidence of helical oscillations \citep[e.g.][]{2016MNRAS.458.4443H, 2017MNRAS.469..639H}.  The phase II results in a decrease in the advance speed and subsequent decrease in aspect ratio.  In our simulations we do not employ such a model, instead relying upon the numerical noise of our simulation and the asymmetrical fluctuations in density created at set-up to deviate the jet as it traverses the non-uniform atmosphere; whilst this is sufficient for low-power jets, it was not sufficient to prevent the higher power runs from forming unrealistically high aspect ratio lobes.

However, it is well known that at least some powerful FR II radio sources are at the same time large, have a stable jet, and a fat lobe (e.g., Pictor A, \cite{2016MNRAS.455.3526H}). A straightforward explanation of the fat lobes may then be jet precession. Other indicators of jet precession have been identified in recent simulations \citep{2020MNRAS.493.3911H, 2020MNRAS.499.5765H, 2022MNRAS.514.5625G} and are frequently seen in powerful radio sources \citep{2019MNRAS.482..240K}.  It appears difficult to explain these findings otherwise.  While it has been widely demonstrated that lighter jets inflate fatter lobes \citep[e.g.][]{2005A&A...431...45K, 2009MNRAS.400.1785G, 2015ApJ...803...48G, 2016ApJ...826...17G,  2021MNRAS.507..175S,  2023MNRAS.523.1104W} and that interaction with the an inhomogeneous ICM can significantly alter the morphology of the lobe \citep[e.g.][]{2011ApJ...728...29W, 2012ApJ...757..136W, 2018MNRAS.479.5544M, 2021AN....342.1140M}, lowering the jet density at constant power would require an increase in the Lorentz factor.  We already use a Lorentz factor of ten, which is towards the upper end of plausible values, as constrained, e.g., by multi-epoch Very Long Baseline Interferometry data.  Hence our jet densities could not plausibly be much lower.  We are therefore left with jet precession as the best explanation for fat radio lobes in galaxy clusters.

\subsection{Resolution study}
An increase in resolution tends to result in a simulation which takes us closer to physical behaviour and closer to convergence \citep[e.g.][]{2003NewAR..47..573K, 2019MNRAS.482.3718P}.  In their resolution study of 3D RMHD jets \cite{2010MNRAS.402....7M} found that higher resolution jets were able to maintain their Lorentz factor better than equivalent lower-resolution runs; whereas \cite{2023MNRAS.523.1104W} showed that the identification of convergence of their models was non-trivial as the factors influencing the length of the jet depended upon the resolution used, with lower resolutions being influenced more by numerical diffusion and higher resolution more by KH instabilities resulting in a turnaround in their jet distance versus time plot, also compare with \cite{2001A&A...380..789K}.  Through the use of stretched grids we have a higher resolution at the centre of our simulation than our previous studies: In comparison with the uniform grids of \citetalias{2014MNRAS.443.1482H,2016MNRAS.461.2025E} and \citetalias{2019MNRAS.490.5807E} the resolution in the x-direction is 3.75 times higher and in the yz-plane is 18.75 times higher at the centre.  In order to judge to what degree resolution influences our models we ran simulations with different numbers of grid points in the x-direction.  In our normal-resolution runs, we have a geometrically stretched zone of 300 cells on either side of the central uniform zone; in our low resolution run we replaced this with a zone of only 150 cells on either side of the central uniform zone, which essentially halved the number of grid points in the x-direction.  In our high resolution run we increased the stretched grid to 300 cells on either side of the central uniform zone, and so essentially doubled the resolution along the x-direction.  We compared the progression of the lobe and the lobe volume/length$^3$ ratio for jet10\_halo30 for all three resolutions (see Figure \ref{2graph_comp1}) and found very similar results, particularly for the two highest resolutions.  Whilst not identical, the fact that the runs all cross in a number of places suggests that the variability between the runs is slight and may be influenced not so much by resolution but by the interpolation of mapping the initial conditions onto a different grid at startup, impacting the exact evolution of the turbulence.  However, our energy analysis (Figure \ref{2graph_comp2}) indicates a more significant difference: the lower resolution run follows a separate path, whereas the higher resolution average lines often fall within the variation indicated by both of their separate lobes, particularly away from the cluster centre.  We conclude from this limited-resolution study that models of evolved lobes are very close to being converged in terms of both dynamics and energetics with models of twice the resolution along the x-axis.

\section{SUMMARY AND CONCLUSIONS}
We created numerical models of realistic cluster atmospheres by utilising the Universal Pressure Profile (UPP) of \cite{2010A&A...517A..92A}, applied a typical temperature profile of a cool core cluster, and ran relativistic helically-magnetised jets into them.  We modelled cluster atmospheres over the range of typically-observed masses, from the group-cluster boundary up to high-mass examples and our jet powers varied from the FR I/FR II boundary up to high-energy values.  In keeping with our previous work, this cluster model also incorporates a multi-scale, tangled magnetic field which is related to the cluster density profile (\cite{2011MNRAS.418.1621H}, \citetalias{2014MNRAS.443.1482H}, \citetalias{2016MNRAS.461.2025E} and \citetalias{2019MNRAS.490.5807E}).

This study has built upon our previous work and improved some aspects of those models, moving closer to realism.  Our use of stretched grids has enabled a much greater resolution along the jet axis and so a smaller and therefore more realistic jet injection cross-sectional area (100x smaller than that used previously).  This has removed the unrealistic plumes of lobe material which were created around the cluster core in our previous models (see \citetalias{2016MNRAS.461.2025E}).  In the current work we have employed relativistic jets with a Lorentz factor of $\gamma = 10$, well within the established range of values observed for real jets, whereas our previous relativistic studies (\citetalias{2016MNRAS.461.2025E} and \citetalias{2019MNRAS.490.5807E}) used a value of $\gamma = 2.7$, which is very much at the lowest boundary of the range of realistic values.  In addition, our use of a helical magnetic field (rather than injecting a purely toroidal field) is another step closer to realism.

In keeping with our previous work, we find that higher power jets inflate lobes with a larger aspect-ratio, progress faster and create larger volume lobes for the same age.  In addition, we again demonstrate that the cluster environment has a significant influence on the evolution of the lobe and we see that atmospheres with a larger mass (measured by $M_{500}$) inhibit the growth of the lobes more and result in a lower aspect ratio.  In our previous work we found that the ratio of shocked energy to lobe energy was a constant, whereas here we find that this increases with jet power and decreases with cluster mass.  Our results for the total synchrotron luminosity follow a very similar pattern to that of our previous work (increases with jet power) although previously we found little variation with atmosphere; here we find that lower power jets are impacted more by the cluster mass, the general trend being that lower mass atmospheres result in greater luminosity.  It was seen that, within the range of parameters studied here, the jet power has a much greater influence on the total synchrotron luminosity than cluster mass.

Synchrotron imaging, maps of the Stokes parameters and charts indicating the direction of the magnetic field are very similar to those we created in our previous work, showing the filamentary structure in a greater level of detail as a result of the higher resolution in the centre of the model; favourable comparisons can also be made with the simulations of other researchers \citep[e.g.][]{1990MNRAS.242..623M, 2001ApJ...557..475T} and with observations \citep[e.g.][]{1988ARA&A..26...93S, 1997MNRAS.288..859H}.   In this current study we have also seen the impact of Doppler boosting on the synchrotron image and created sidedness plots which resemble those derived from observations of 3C 31.

Our simulations are the first to use the UPP in a numerical simulation of jet feedback and the resultant density and synchrotron images of the lobes compare very well with images of radio galaxies from observations.  Using the same jet parameters, we created an equivalent $\beta$-profile atmosphere and conclude that whilst the dynamics of the UPP and $\beta$-profile are very similar (particularly at large distances from the core), the greatest differences lie in the energetics where significant variations can be seen in the distribution of energy between the lobe and shock regions and also between the magnetic and thermal energies of the lobes.  There is also a significant morphological difference in that, for the parameters chosen for comparison, the UPP profile enables more lobe material to escape the centre of the cluster and form two distinct lobes on either side of the injection cylinder, whereas the $\beta$-model does not provide sufficient buoyancy to achieve this.  In real clusters, cold material surrounding the cluster core impedes the progression of jets; in addition, rising bubbles lift material away from the cluster core which would otherwise have been accreted onto the SMBH and so starve it of material from which to form the jets.  This process would result in the jets carrying less energy and so reduce the energy carried away by the bubbles.  Such thermostatic control of an AGN is a key element of the feedback loop of real clusters \citep{2007ARA&A..45..117M, 2012NJPh...14e5023M, 2016ApJ...829...90Y}, which regulates the power of the jets; our models have a constant power and so we do not model this aspect of AGN feedback.

Our simulations have highlighted the role of instabilities in the progression of the jet.  Where the jet is exposed to the ambient medium through the buoyancy of the lobe having removed it from the core of the AGN, we observe the growth of KH instabilities; although we believe these have been exaggerated by our jet injection boundary condition through the formation of a slower-moving sheath around the jet.  While short-wavelength modes of the KH instability on the surface of the jets are suppressed through resolution effects, this happens in a similar way for all simulations. The preferential suppression of the KH instability in the high power jets is hence likely mainly caused by the shielding by the low-density lobes. This drives the latter to high aspect ratios, greater than what is generally seen in observations.  Our work highlights the limitations of such numerical simulations and points to the need to introduce a mechanism to increase the jet working surface at extended distances, this can be done with the introduction of a `wobble' such as that used by \cite{2019MNRAS.482.3718P, 2022MNRAS.510.2084P}.  Lobes wider than those seen in our simulations can easily be produced using a precession mechanism, which can be caused by a supermassive secondary or a misaligned accretion disc \citep{2019MNRAS.482..240K, 2020MNRAS.493.3911H, 2020MNRAS.499.5765H, 2023MNRAS.521.2593H}. The latter can, however, also be related to a binary supermassive black hole \citep{2022MNRAS.509.5608N}.

Where our images depart from current observations, this is due to the lack of non-thermal particles in our model; this means that shock acceleration cannot be simulated and so the knots and hot spots normally seen are absent.  Ground-breaking work is being carried out by others on this area (see \cite{2021MNRAS.505.2267M} and \cite{2021MNRAS.508.5239Y}) which have been successful in reproducing the features of DSA.  Similarly, our models do not take into account radiative losses or spectral ageing, unlike the work of \cite{2021MNRAS.508.5239Y}.

Future papers will use this UPP atmosphere model to probe the synthetic X-ray emission from the hot gas at the centre of the cluster and also to further investigate synthetic radio polarisation images of these clusters, extending our previous work to these more realistic models.

\section*{Acknowledgements}

MJH acknowledges support from the UK STFC [ST/V000624/1].  This work has made use of the University of Hertfordshire Science and Technology Research Institute high performance computing facility (\url{https://uhhpc.herts.ac.uk/}).  We also thank an anonymous referee for a careful and constructive reading of this paper which led to significant improvements upon the original text.

\section*{Data Availability}

Data available upon request.


\bibliographystyle{mnras}
\bibliography{References} 



\appendix

\section{Cluster magnetic fields}
\subsection{Theory of cluster magnetic fields}\label{obs_mag_fields}
The radial scaling of the cluster magnetic field strength with density distribution was first put forward by \cite{1980ApJ...241..925J}.  \cite{2011MNRAS.410.2446K}, from theoretical work on the viscous heating of the ICM suggest a magnetic field which scales as $B\propto n_e^{1/2}$.  This result is also recovered by \cite{2010A&A...513A..30B} from rotation measure (RM) observations of the Coma cluster.  This relation is assumed by many authors \citep[e.g.][]{2015ApJ...800...60M,2015Natur.523...59M} and can be expressed as
\begin{equation}
\frac{B^2}{8\pi}=\eta \frac{3}{2} nk_B T\label{magden_eqn}
\end{equation}
where $\eta$ is the energy density ratio between the magnetic field and the thermal energy density.  If the magnetic field strength decreases as the square root of the thermal electron density, then the gas must remain in overall hydrostatic equilibrium since the magnetic energy density decreases in proportion with the gas energy density.  Through their work in comparing models with observations, \cite{2016A&A...596A..22B} find that the magnetic energy density is two orders of magnitude smaller than the thermal energy with $\eta$ (in Equation \ref{magden_eqn}.) falling in the range $5-\num{7.5e-3}$; this means that the magnetic fields in clusters are considered to be dynamically unimportant.  In this study we will employ a magnetized cluster with a dynamically unimportant magnetic energy density scaling with the gas energy density, producing a distribution dependent upon the model chosen for the density distribution.

\subsection{Creating the model cluster magnetic field}\label{model_magfield}
We define a vector potential of the form $\tilde{A}(k)=A(k)e^{i\theta (k)}$, where $k$ is the wave-vector ($k^2=k_x^2+k_y^2+k_z^2$), $i$ is the unitary complex number and $A$ and $\theta$ are the vector's amplitudes and phases (compare \cite{1991MNRAS.253..147T}).  Values for $\theta (k)$ are then drawn from a uniform random distribution, whereas values for $\textbf{A}(\textbf{k})$ are drawn from a Rayleigh probability distribution, given in polar form as:
\begin{equation}
P\left(A,\theta \right)\mathrm{d} A \mathrm{d} \theta =\frac{A}{2\pi|A_k|^2}
\exp \left(-\frac{A^2}{2|A_k|^2}\right)\,\mathrm{d} A\mathrm{d}\theta,
\end{equation}
where $|A_k|^2 \propto k^{-\zeta/2}$.  It is then possible to transform $\tilde{A}(k)$ to real space by employing an inverse Fourier transform.  The three vector potential components are then used to provide the magnetic field strength (i.e. $B(x)=\nabla \times A(x)$) resulting in an isotropic divergence-free tangled magnetic field:
\begin{equation}
|B_k|^2=C_n^2k^{-n}
\end{equation}
where $n=\zeta -2$ and $C_n^2$ is the power spectrum normalisation.  \cite{2011MNRAS.418.1621H} then employed a Kolmogorov-like 3D turbulent slope $n=11/3$, which is based upon both theoretical studies of how CMFs evolve \citep{2009A&A...504...33V} and observations of pressure fluctuations in the Coma cluster \citep{2004A&A...426..387S}.

\begin{figure}
\includegraphics[width=0.5\textwidth]{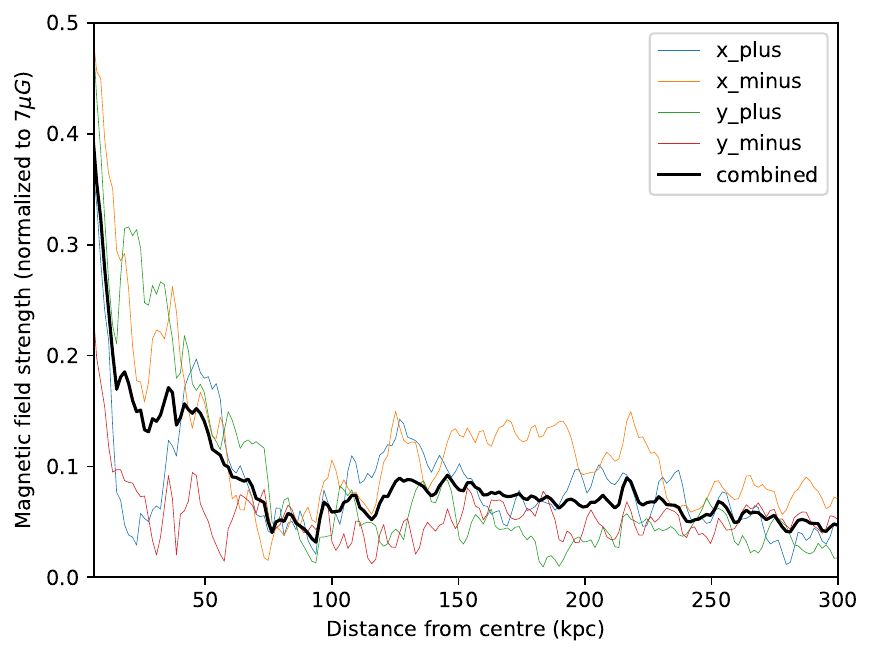}
\caption{Variation of magnetic field strength with distance from the cluster centre at $t=0$ for a UPP atmosphere with $M_{500}=3\times\num{e14}\,\text{h}^{-1}_{70}M_{\odot}$.  Individual colours are along separate radii from the centre, the black line is the average of these.  All values are normalized to $7\mu$G.}
\label{magnetic graph}
\end{figure}

\cite{2011MNRAS.418.1621H} allowed the magnetized plasma to relax for a period of time ($118\,$Myr) before injecting their jets to ensure that the atmosphere was stable.  However, here we follow the method introduced in \citetalias{2014MNRAS.443.1482H} and cut the scale off below $\sim3$ pixels to avoid injecting high-spatial-frequency structure into the simulations (this is done by removing the wavevectors greater than $1/3$ of the Nyquist frequency in Fourier space).  These high frequencies would have been damped out by numerical diffusion within a short space of time at the start of the run.  This, therefore, provides a more useful baseline at $t=0$ such that jets can be introduced into a near-stable environment from the very start of the run, maximising valuable computer-processing time.
For consistency we will follow \citetalias{2016MNRAS.461.2025E} in setting the field strength to $0.7\,$nT when the the density (in simulation units) is unity (a realistic value derived from observations of the Coma cluster by \cite{1995A&A...302..680F}).  The introduction of the tangled magnetic field in our model creates a small amount of turbulence in the initially static ICM due to magnetic tension; the ratio of kinetic energy to thermal energy as a result of this was measured to be $\sim\num{e-6}$ and so our cluster models are, to a good approximation, relaxed.  Our methodology contrasts with the work of some other researchers who deliberately set out to investigate dynamic cluster environments \citep[e.g.][]{2018MNRAS.481.2878E, 2021MNRAS.503.1327E}.  We will follow \citetalias{2016MNRAS.461.2025E} in increasing the magnetic field energy density in proportion to the gas energy density (Equation \ref{magden_eqn}) and so $B \propto \sqrt{n_e}$.  The variation of the magnetic field strength with distance from the cluster centre can be seen in Figure \ref{magnetic graph} where its variable and multi-scaled nature is evident.  Plasma $\beta$ values were calculated for the atmosphere shown in Figure \ref{magnetic graph} with typical values falling in the range $\num{e2} - \num{e3}$; these coincide with values found in the literature \citep[e.g.][]{2016MNRAS.459.2701M, 2023Galax..11...73B} and indicate that the magnetic field of our cluster models is not dynamically important.

\section{Simulated Polarimetry}\label{stokes_theory}
The polarized emission from the lobe can be characterised by means of the Stokes parameters.  Firstly, relativistic aberration needs to be accounted for; this is where the angle between the light ray and the velocity direction in the observer's frame of reference $\theta_o$ will be different to that of the object's frame $\theta_s$ when moving at relativistic speeds.  The formula is \citep{1905AnP...322..891E}
\begin{equation}
\cos \theta_o = \frac{\cos \theta_s -\beta}{1 - \beta \cos \theta_s} 
\end{equation}
where $\beta$ is the velocity in units of the speed of light.  Therefore, we define here $B_x$ and $B_y$ as the components perpendicular to the aberration-corrected projection axis.  The Stokes $I$ (total intensity) and Stokes $Q$ and $U$ (polarized intensities) parameters are then calculated (in simulation units) by summing the following relations over the emission volume
\begin{equation}
j_I=p\left(B_x^2+B_y^2\right)^{\frac{1}{2}(\alpha -1)}(B_x^2+B_y^2)D^{3+\alpha}\label{RI}
\end{equation}
\begin{equation}
j_Q=\mu p\left(B_x^2+B_y^2\right)^{\frac{1}{2}(\alpha -1)}(B_x^2-B_y^2)D^{3+\alpha}\label{RQ}
\end{equation}
\begin{equation}
j_U=\mu p\left(B_x^2+B_y^2\right)^{\frac{1}{2}(\alpha -1)}(2B_xB_y)D^{3+\alpha}\label{RU}
\end{equation}
where $p$ is the local thermal pressure, $\alpha$ is the power-law synchrotron spectral index, which is taken to be $\alpha=0.5$ corresponding to an electron energy index $p=2$ and $\mu$ is the maximum fractional polarization for a given spectral index: for $\alpha = 0.5$: $\mu=(\alpha +1)/(\alpha +5/3)=0.69$.  D is the Doppler factor, given by
\begin{equation}
D=\frac{1}{\gamma(1-\beta \cos(\theta))},
\end{equation}
where $\gamma$ is the Lorentz factor and $\theta$ is the angle between the projection vector and the velocity vector of the cell.  The scaling factor to convert from simulation units to SI units is given as $j_0=\num{3.133e31}\,$WHz$^{-1}$sr$^{-1}$, where a constant observing frequency of $151\,$MHz has been assumed.  The Stokes parameters can then be used to calculate the linear polarisation $\Pi$
\begin{equation}
\Pi =\frac{\sqrt{Q^2+U^2}}{I}.\label{linpol2}
\end{equation}
Given the magnetic field is perpendicular to the electric field, then the observable direction of the magnetic field (projected onto the sky) $\theta_B$ is given by
\begin{equation}
\theta_B=\frac{1}{2}\arctan\left(\frac{U}{Q}\right) + \frac{\pi}{2}.
\end{equation}


\bsp	
\label{lastpage}
\end{document}